\documentclass[11pt,a4paper]{article}

\usepackage[a4paper,margin=1in]{geometry}
\usepackage{indentfirst}

\usepackage{amsmath,amssymb}

\usepackage{graphicx}
\usepackage{subcaption}

\usepackage{dcolumn}

\usepackage{cite}

\usepackage{xcolor}
\usepackage{hyperref}

\title{Unimodular Diffusion and Interacting Vacuum Cosmology}

\author{
Gopal Kashyap$^{1}$\thanks{Email: gplkumar87@gmail.com} and
Naveen K. Singh$^{2}$\thanks{Email: naveen.nkumars@gmail.com} \\
\vspace{0.1cm}
$^{1}$Department of Physics, School of Advanced Sciences,\\
Vellore Institute of Technology, Vellore 632014, Tamil Nadu, India \\
$^{2}$Sir P. T. Sarvajanik College of Science (Autonomous),\\
Surat 395001, Gujarat, India
}

\date{}

\begin{document}

\maketitle

\begin{abstract}
We investigate the correspondence between unimodular diffusion cosmology and interacting dark sector models at the background and linear perturbation levels. In the diffusion framework, the effective cosmological constant becomes time dependent, $\Lambda(t)$, sourced by a diffusion current. We show that at the background level this framework can be mapped onto interacting dark energy models with $w=-1$ and energy transfer $Q$. Using two common parameterizations, $Q = \xi H \rho_{\rm de}$ and $Q = \xi H \rho_{\rm dm}$, and data from supernovae, DESI BAO, cosmic chronometers, and CMB distance priors, we find $\xi = -0.0197 \pm 0.0076$ for the vacuum-coupled case, while the matter-coupled case gives a best-fit $\xi = 0.0018$ with comparable goodness of fit.
At the level of linear perturbations, however, the diffusion framework is consistent only with interacting vacuum models having homogeneous energy transfer ($Q \propto \rho_{\rm de}$ with $\delta Q=0$), thereby breaking the degeneracy with more general interacting dark energy scenarios. Including redshift-space distortion data, we obtain $\xi = -0.0147 \pm 0.0075$, consistent with $\Lambda$CDM ($\xi=0$) at $2\sigma$. The inferred clustering amplitude is $S_8 = 0.782 \pm 0.026$ for the diffusion model, compared to $S_8 = 0.77 \pm 0.025$ for $\Lambda$CDM under the same dataset, indicating a modest but non-negligible impact on structure growth.
\end{abstract}

\section{Introduction}

Over the past two decades, a wide range of observations have established that the universe is undergoing a phase of accelerated expansion \cite{Riess,Perlmutter}. Within the standard cosmological model, this phenomenon is attributed to a dark energy component with negative pressure, most simply realized as a cosmological constant $\Lambda$. The resulting $\Lambda$CDM model provides an excellent fit to diverse datasets, including cosmic microwave background (CMB) anisotropies \cite{Planck:2018vyg}, large-scale structure observations \cite{BOSS:2016}, and measurements of the late-time expansion history.

In this framework, the energy content of the universe is dominated by dark energy ($\Omega_{\rm de} \simeq 0.68$) and cold dark matter ($\Omega_{\rm dm} \simeq 0.27$), with a subdominant baryonic component \cite{Planck:2018vyg}. The equation-of-state parameter of dark energy is consistent with a cosmological constant, $w \approx -1$, within current observational uncertainties. Despite its success, the $\Lambda$CDM model faces several conceptual and observational challenges, including the cosmological constant problem \cite{Weinberg:1988cp}, the coincidence problem \cite{Copeland:2006wr}, and persistent tensions between independent cosmological probes. In particular, the discrepancy between early and late-time determinations of the Hubble constant $H_0$ \cite{Riess:2020fzl,H0LiCOW:2018tyj}, as well as recent indications from DESI data of a possible deviation from $w=-1$ \cite{DESI:2024mwx,DESI:2025zgx}, motivate the exploration of extensions beyond the standard model.

Among the proposed alternatives, interacting dark sector models have received considerable attention \cite{Luca_2000,ZIMDAHL2001,PhysRevD.74.023501}. In these models, dark matter and dark energy exchange energy through a phenomenological coupling term $Q$, modifying the continuity equations of the individual components while preserving conservation of the total energy-momentum tensor. A particularly well-motivated subclass corresponds to interacting vacuum energy models with $w=-1$, in which the vacuum energy exchanges energy with dark matter \cite{Wands:2012vg,Borges:2020cbh,Wang:2014xca,Wang:2015wga,Kaeonikhom:2022jyu}. Such interactions have been explored as possible mechanisms to alleviate cosmological tensions and address the coincidence problem \cite{Hu:2006ar,Sadjadi:2006qp,Berger:2006db,Murgia:2016ccp,Yang_2018,Guo:2021rrz,DIVALENTINO_2020,PhysRevD.105.123506,Guin_2025}. Recent studies have also explored interacting dark sector models in the context of the latest cosmological observations, including DESI measurements \cite{Li:2024qso,Pan:2025qwy,Silva:2025hxw}. For instance, in Ref.~\cite{Guo:2021rrz} the interaction term is taken to be proportional to the dark matter energy density, while in Ref.~\cite{Pandey:2019plg} the decay of dark matter into dark radiation has been discussed as a possible mechanism to address both the Hubble tension and the $S_8$ tension. However, the absence of a fundamental microphysical origin for the interaction term remains a significant limitation, and different choices of $Q$ are largely phenomenological.

A conceptually distinct framework arises in unimodular gravity, where the determinant of the metric is fixed and the trace of Einstein’s equations is removed. In this formulation, the cosmological constant emerges as an integration constant rather than a fundamental parameter of the action \cite{Buchmuller:1988wx,Unruh:1988in}. Various aspects and extensions of unimodular gravity have been extensively explored in the literature (see, e.g., Refs.~\cite{Jain:2011jc,Jain:2012gc,Gao2014,Nojiri:2016ppu,Bamba:2016wjm,Costantini:2022nof,Cho:2014taa,Singh:2012sx,Odintsov:2016imq,Cedeno2021,Agrawal2023,universe9110469,Das_2023}).

An important consequence of this framework is the emergence of modified conservation laws, arising from the structure of the field equations, which allow for an effective energy exchange between matter and a dynamical vacuum component \cite{PhysRevLett.118.021102}. Diffusion models provide a natural realization of such energy exchange within unimodular cosmology \cite{PhysRevD.102.023508,cedeno_2021,PhysRevD.108.043524}.

In this framework, the modified conservation equation for dark matter can be written as
\begin{equation}
\dot{\rho}_{\rm dm} + 3H(\rho_{\rm dm}+p_{\rm dm}) = -\dot{P},
\end{equation}
where $P(t)$ is a diffusion function encoding the energy flow. This relation leads to an effective interaction term $Q = \dot{P}$, allowing the diffusion framework to be formally mapped to interacting dark sector models. However, this correspondence is restricted, diffusion cosmology can be mapped onto interacting models only in the case of vacuum-like dark energy with $w=-1$, even at the level of homogeneous expansion.

This distinction raises a central question ``to what extent can diffusion cosmology be observationally distinguished from phenomenological interacting dark sector models?'' In this work, we address this issue by systematically analysing the correspondence between these frameworks at both the background and linear perturbation levels. We show that, although a mapping exists at the background level under the restriction $w=-1$, this correspondence does not persist in general once perturbations are taken into account. In particular, the diffusion framework is consistent only with a restricted class of interacting vacuum models characterized by homogeneous energy transfer, a feature known to play a crucial role in the perturbative stability and growth of structure in such models \cite{Wands:2012vg,Borges:2020cbh}. This breaks the degeneracy with more general interacting scenarios and provides a physically motivated and observationally testable criterion to distinguish between the two approaches. It is important to emphasize that this mapping is not fundamental, but arises from the modified conservation laws in unimodular gravity. In particular, it does not correspond to a fully covariant interacting dark energy model within standard General Relativity, and is therefore restricted both dynamically and at the level of perturbations.

To test these predictions, we constrain the model using current cosmological observations, including the Pantheon+ Type Ia supernova sample, cosmic chronometer measurements of the Hubble parameter $H(z)$, baryon acoustic oscillation (BAO) data from DESI DR2, redshift-space distortion (RSD) measurements, and CMB distance priors. These datasets allow us to probe both the background expansion history and the growth of cosmic structure. Observational constraints on interacting vacuum energy models using similar datasets have been explored in Refs.~\cite{Wang:2014xca,Wang:2015wga,Kaeonikhom:2022jyu}.

The paper is organized as follows. In Section~\ref{sec:int} we review interacting dark sector models. Section~\ref{sec:diff-int-map} presents the diffusion framework and its mapping to interacting scenarios. The background cosmological evolution is discussed in Section~\ref{sec:background}, while the datasets and methodology are described in Section~\ref{sec:data_method}. Observational constraints and model comparison are presented in Section~\ref{sec:background_constraints}. The analysis of linear perturbations is given in Section~\ref{sec:perturbations}, and conclusions are summarized in Section~\ref{sec:con}.

\section{Review of Interacting Dark Sector Models}
\label{sec:int}

Interacting dark sector models allow for an exchange of energy between dark matter and dark energy through a phenomenological interaction term $Q$.
In such case the individual components are not conserved separately,
although the total energy-momentum tensor remains conserved~\cite{Luca_2000, ZIMDAHL2001}.

Assuming pressureless dark matter ($p_{\rm dm}=0$) and dark energy with
equation of state
\begin{equation}
p_{de} = \omega_{de}\rho_{de},
\end{equation}
the continuity equations for the dark sector components can be written as
\begin{align}
\dot{\rho}_{dm} + 3H\rho_{ dm} &= -Q, \label{DMEQN}\\
\dot{\rho}_{de} + 3H(1+\omega_{de})\rho_{de} &= Q .
\label{DEEQN}
\end{align}

With this convention, $Q>0$ corresponds to energy transfer from dark matter
to dark energy, while $Q<0$ represents the reverse process. Since the
microscopic origin of the dark sector interaction is unknown, the functional
form of $Q$ is typically introduced phenomenologically. Two commonly studied
interaction forms are~\cite{He_2008}
\begin{equation}
Q = \xi H\rho_{ dm}, \qquad
Q = \xi H\rho_{de},
\end{equation}
where $\xi$ is a dimensionless coupling parameter.

For the interaction $Q=\xi H\rho_{ dm}$, Eq.~(\ref{DMEQN}) can be solved
analytically, yielding
\begin{equation}
\rho_{ dm}(a) = \rho_{ dm0} a^{-3-\xi}.
\end{equation}
Thus the interaction modifies the standard dilution law of matter.
For $\xi=0$ the standard $\Lambda$CDM scaling $\rho_{ dm}\propto a^{-3}$
is recovered, while negative (positive) values of $\xi$ lead to a slower
(faster) dilution of dark matter.

We now consider the interaction $Q=\xi H\rho_{de}$. For illustration we
adopt a time-dependent equation-of-state parameter of the form
\begin{equation}
\omega_{de}(a) = -1 + \alpha a + \beta a^2 ,
\end{equation}
which allows small deviations from a cosmological constant.

Using Eq.~(\ref{DEEQN}) and transforming derivatives to the scale factor
via $d/dt = aH\, d/da$, the dark energy density evolves as
\begin{equation}
\rho_{de}(a)
=
\rho_{de0}\,
a^{\xi}
\exp\!\left[-3\left(\alpha(a-1)+\frac{\beta}{2}(a^2-1)\right)\right].
\end{equation}

Substituting this result into Eq.~(\ref{DMEQN}), the dark matter density
satisfies
\begin{align}
\frac{d\rho_{ dm}}{da}+\frac{3}{a}\rho_{ dm}=-&\xi \,\rho_{de0}\,a^{\xi-1} \times \exp\!\left[-3\left(\alpha(a-1)+\frac{\beta}{2}(a^2-1)\right)\right].
\end{align}
Solving this first-order equation gives
\begin{align}
\rho_{\rm dm}(a) = & a^{-3}\Big[\rho_{\rm dm0}- \xi\rho_{de0}\int_{1}^{a} a'^{\xi+2}
 \times \exp\!\Big(-3\big(\alpha(a'-1) + \tfrac{\beta}{2}(a'^2-1)\big)\Big)\, da' \Big].
\end{align}
This expression shows explicitly that the dark matter density receives
a contribution sourced by the evolving dark energy component.

\begin{figure*}[t]
\centering
\includegraphics[width=0.48\linewidth]{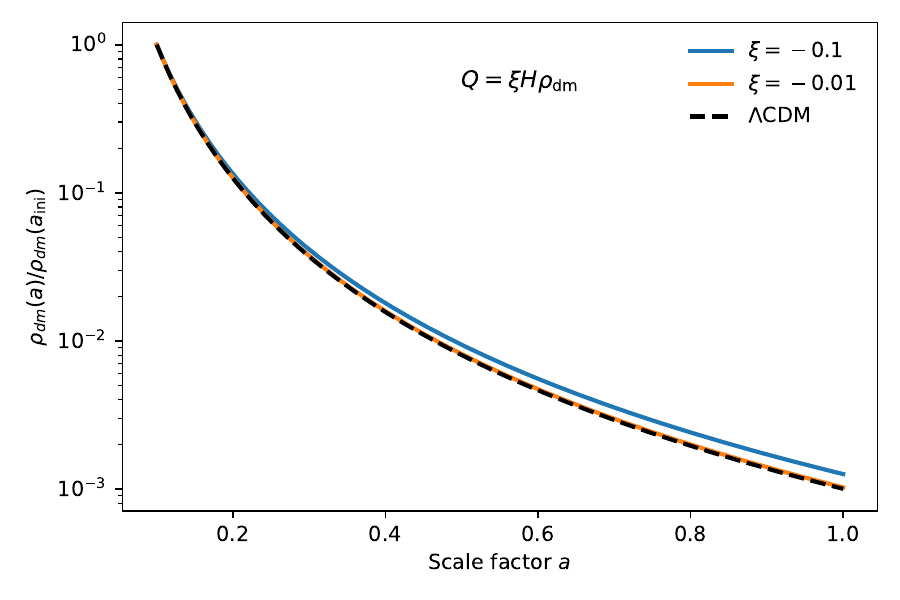}
\hfill
\includegraphics[width=0.48\linewidth]{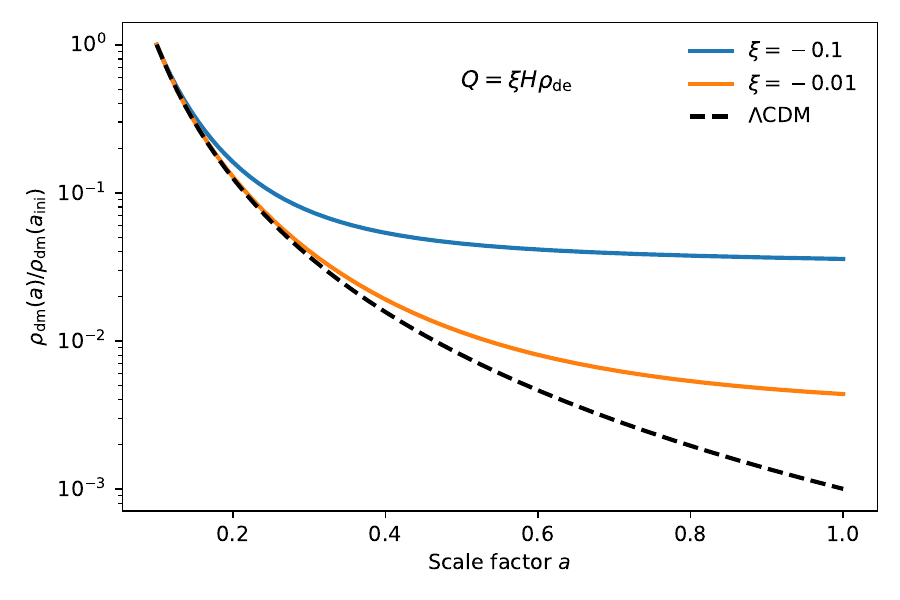}
\caption{Evolution of the normalized dark matter density for interacting
dark sector models. \textit{Left}: interaction $Q=\xi H\rho_{\rm dm}$ showing
the modified dilution law $\rho_{\rm dm}\propto a^{-3-\xi}$. 
\textit{Right}: interaction $Q=\xi H\rho_{de}$ where the matter density
evolution depends on the dark energy dynamics.}
\label{fig_density}
\end{figure*}

The different interaction prescriptions lead to distinct modifications of
the matter density evolution. For the coupling $Q=\xi H\rho_{ dm}$ the
matter density follows a simple power-law scaling $a^{-3-\xi}$, whereas
for $Q=\xi H\rho_{de}$ the evolution depends on the dark energy
dynamics through the above integral solution. The resulting behavior of $\rho_{ dm}(a)$ for the considered interaction models is shown in Fig.~\ref{fig_density}. For the numerical analysis we choose $\alpha=0.01$ and $\beta=0.001$, which correspond to
small departures from the $\Lambda$CDM limit ($w=-1$) and therefore serve
only to demonstrate the qualitative effect of the interaction on the matter
density evolution. The evolution of the normalized dark matter density for the case $(Q=\xi H\rho_{dm})$ deviates relatively little from the standard $\Lambda$CDM behavior, and the matter density evolves close to the usual $a^{-3}$ scaling. In contrast, for $(Q=\xi H\rho_{ de})$, the interaction significantly modifies the dilution rate of dark matter, leading to a noticeably slower decrease of $\rho_{dm}$ for negative values of $\xi$. The effect becomes stronger as $\xi$ increases, illustrating how the interaction parameter controls the energy transfer in the dark sector.

These interacting dark sector models provide a useful phenomenological
framework for describing energy exchange in the dark sector. In the next
section we show that the diffusion model arising from unimodular gravity
can be mapped onto a specific class of interacting vacuum models,
leading to an exact degeneracy at the background level.

\section{Diffusion dynamics and mapping to interacting running vacuum models}
\label{sec:diff-int-map}

In unimodular gravity, the conservation of the energy--momentum tensor is modified by the presence of a diffusion current (see, e.g., Refs.~\cite{PhysRevD.102.023508,cedeno_2021} for a detailed discussion). The resulting generalized conservation equation can be written as,
\begin{equation}
\nabla_{\mu}T^{\mu\nu}=J^{\nu},
\end{equation}
where $J^{\nu}$ describes the diffusion of energy between matter and
the vacuum sector. For a homogeneous and isotropic FLRW universe the
time component of this equation leads to a modified continuity equation
for dark matter,
\begin{equation}
\dot{\rho}_{\rm dm}+3H\rho_{\rm dm}=-\dot P ,
\label{dm_diff}
\end{equation}
where $P(t)$ denotes the diffusion function.

In unimodular gravity the Lagrange multiplier $\lambda$ appearing
in the field equations behaves as a dynamical vacuum contribution.
Its evolution is related to the diffusion process through
\begin{equation}
\dot{\lambda}=8\pi G\,\dot P .
\end{equation}
It is therefore convenient to define an effective vacuum energy
density
\begin{equation}
\rho_{\rm de} \equiv \frac{\lambda}{8\pi G}.
\end{equation}
Using this definition we obtain
\begin{equation}
\dot{\rho}_{\rm de}=\dot P .
\label{de_diff_eq}
\end{equation}

Eqs.~(\ref{dm_diff}) and (\ref{de_diff_eq}) can be written in the
same form as the continuity equations of interacting dark sector
models,
\begin{align}
\dot{\rho}_{\rm dm}+3H\rho_{\rm dm} &= -Q, \\
\dot{\rho}_{\rm de}+3H(1+w_{\rm de})\rho_{\rm de} &= Q ,
\end{align}
by identifying the interaction term
\begin{equation}
Q=\dot P .
\end{equation}
Comparison with Eq.~(\ref{de_diff_eq}) shows that consistency
requires
\begin{equation}
w_{\rm de}=-1 ,
\end{equation}
indicating that the effective dark energy component behaves as a
vacuum energy with equation of state $w=-1$.

The relation $Q=\dot P$ therefore establishes a direct connection
between the diffusion process and interacting dark sector models.
However, the diffusion framework does not uniquely determine the
functional dependence of the interaction term on the dark sector
densities. In phenomenological studies it is common to parameterize
the interaction rate as proportional to the Hubble expansion rate
and one of the dark sector densities. Following the discussion in
Sec.~\ref{sec:int}, we therefore consider the two widely studied
interaction models
\begin{equation}
Q=\xi H\rho_{\rm de}, \qquad Q=\xi H\rho_{\rm dm},
\end{equation}
where $\xi$ is a dimensionless coupling parameter.

These parameterizations can be interpreted as different
phenomenological aspects of the diffusion source term
$Q=\dot P$. Consequently, once the interaction term is specified,
the unimodular diffusion model can be mapped onto interacting dark
sector models at the level of the homogeneous background
evolution. The resulting cosmological expansion histories are
therefore degenerate and cannot be distinguished using background
observables alone.

This degeneracy can be broken at the level of cosmological perturbations.
While the background dynamics depend only on the effective interaction term
$Q$, the perturbation equations retain sensitivity to the physical origin of
the interaction. As a result, diffusion-driven vacuum dynamics can in
principle be distinguished from phenomenological interacting vacuum models
through large-scale structure, anisotropy, and growth-rate observations.
\section{Background Dynamics and Observables}
\label{sec:background}

We consider spatially flat Friedmann-Lemaitre-Robertson-Walker (FLRW)
universe composed of radiation ($r$), baryons ($b$), dark matter
($\mathrm{dm}$), and dark energy ($\mathrm{de}$). Radiation and baryons are
assumed to be separately conserved, while an interaction is allowed between
dark matter and dark energy. The corresponding continuity equations are
\begin{align}
\dot{\rho}_{r} + 4H \rho_{r} &= 0, \\
\dot{\rho}_{b} + 3H \rho_{b} &= 0, \\
\dot{\rho}_{\mathrm{dm}} + 3H \rho_{\mathrm{dm}} &= -Q, \\
\dot{\rho}_{\mathrm{de}} + 3H(1+w_{\mathrm{de}})\rho_{\mathrm{de}} &= Q ,
\end{align}
where $Q$ denotes the energy transfer between the dark sector components.
Following the mapping discussed in Sec.~\ref{sec:diff-int-map}, we impose
$w_{\mathrm{de}}=-1$ and identify the interaction parameter of the interacting
dark sector model with the diffusion coupling parameter.

The Friedmann equations retain their standard form,
\begin{align}
H^2 &= \frac{8\pi G}{3}
\left(
\rho_{r} + \rho_{b} + \rho_{\mathrm{dm}} + \rho_{\mathrm{de}}
\right), \label{friedmann_full} \\
\dot{H} &= -4\pi G
\left(
\rho_{r} + \rho_{b} + \rho_{\mathrm{dm}} + \rho_{\mathrm{de}} +
P_{\mathrm{de}} + P_r
\right).
\end{align}

For vacuum-like dark energy ($w_{\mathrm{de}}=-1$), the dark energy pressure
satisfies $P_{\mathrm{de}}=-\rho_{\mathrm{de}}$. Substituting this relation
into the Raychaudhuri equation gives
\begin{equation}
\dot{H} = -4\pi G
\left(
\rho_{\mathrm{dm}} + \rho_{b} + \frac{4}{3}\rho_{r}
\right).
\label{hdot_full}
\end{equation}

Although the interaction term $Q$ does not appear explicitly in
Eq.~\eqref{hdot_full}, its effects are encoded in the evolution of the dark
matter and dark energy densities through the modified continuity equations.

\subsection{Deceleration parameter and effective equation of state}
\label{sec:q_weff}

The deceleration parameter is defined as
\begin{equation}
q \equiv -1 - \frac{\dot{H}}{H^2}.
\end{equation}
Using the Friedmann and Raychaudhuri equations,
Eqs.~\eqref{friedmann_full} and \eqref{hdot_full}, this can be expressed in
terms of the total energy density and pressure as
\begin{equation}
q=\frac{1}{2}\left(1 + 3\,\frac{p_{\rm tot}}{\rho_{\rm tot}}\right)
= \frac{1}{2}\left(1 + 3 w_{\rm eff}\right),
\label{q_weff_def}
\end{equation}
where we define the effective equation of state
\begin{equation}
w_{\rm eff} \equiv \frac{p_{\rm tot}}{\rho_{\rm tot}} .
\end{equation}

For a universe containing radiation, baryons, dark matter, and vacuum energy
with $w_{\mathrm{de}}=-1$, the deceleration parameter can be written as
\begin{equation}
q=-1+\frac{3}{2}
\frac{
\rho_{\mathrm{dm}} + \rho_{b} + \frac{4}{3}\rho_{r}
}{
\rho_{r} + \rho_{b} + \rho_{\mathrm{dm}} + \rho_{\mathrm{de}}
}.
\label{qgeneral_full}
\end{equation}

Equivalently, in terms of the density parameters,
\begin{equation}
q=-1+\frac{3}{2}\left(
\Omega_{\mathrm{dm}} + \Omega_{b} + \frac{4}{3}\Omega_{r}
\right).
\end{equation}

The functional forms of Eqs.~\eqref{q_weff_def} and \eqref{qgeneral_full}
follow directly from the background field equations and are therefore
independent of the explicit form of the dark sector interaction. However,
different interaction terms modify the redshift evolution of the individual
energy densities, leading to distinct histories for $w_{\rm eff}(z)$ and
$q(z)$.

\section{Data and Methodology}
\label{sec:data_method}

We now test the interacting dark sector models against cosmological data. As shown in Sec.~\ref{sec:diff-int-map}, these models have the same background expansion history as unimodular diffusion, making them observationally equivalent at the background level.
\subsection{Datasets}

The analysis incorporates the following observational datasets:

\begin{itemize}

\item \textbf{Cosmic chronometers:} 
We use 32 measurements of the Hubble expansion rate obtained from the cosmic chronometer (CC) method. See Table \ref{tab:Hz} and the references. This approach estimates $H(z)$ directly from the differential age evolution of the galaxies, using the relation $H(z) = -\frac{1}{1+z}\frac{dz}{dt}$~\cite{Jimenez_2002}. Since it relies only on the relative ages of galaxies, the CC technique provides a model independent probe of the expansion history without assuming a specific cosmological model.

\item \textbf{Type Ia Supernovae (Pantheon+):}  
We use the Pantheon+ compilation of Type Ia Supernovae (SNIa)\footnote{\url{https://github.com/CobayaSampler/sn_data}} \cite{Scolnic_2022}, which contains 1701 light curve measurements corresponding to 1550 unique 
spectroscopically confirmed SNIa spanning the redshift range 
$0.001 < z < 2.26$. The additional light curves arise from repeated 
observations of certain objects, including supernova siblings, multiple 
supernovae occurring in the same host galaxy, as well as SNIa located in 
Cepheid calibrated host galaxies that are used to anchor the distance scale.

\item \textbf{Baryon Acoustic Oscillations (DESI DR2):}  
We further include baryon acoustic oscillation (BAO) measurements from the second data release of the Dark Energy Spectroscopic Instrument (DESI DR2)\footnote{\url{https://github.com/CobayaSampler/bao_data/tree/master/desi_bao_dr2}} \cite{desi_dr2}. BAO observations provide a standard ruler set by the sound horizon at the baryon drag epoch and probe the expansion history through measurements of the comoving angular diameter distance and the Hubble parameter at different redshifts.

\item \textbf{CMB distance priors:} For the cosmic microwave background (CMB) we use distance priors derived from the 
\textit{Planck} 2018 observations \cite{Chen_2019}. This approach efficiently constrains background cosmology while avoiding a full Boltzmann analysis.
\end{itemize}
\begin{table}[h!]
\centering
\caption{$H(z)$ measurements with $1\sigma$ uncertainties used in this work.}
\label{tab:Hz}
\setlength{\tabcolsep}{2.5pt}   % tighten column spacing
\renewcommand{\arraystretch}{0.95} % tighten row spacing
\footnotesize  % slightly smaller font (acceptable)

\begin{tabular}{c c c c c c c c}
\hline\hline
$z$ & $H$ & $\sigma$ & Ref. & $z$ & $H$ & $\sigma$ & Ref.\\
\hline
0.07 & 69.0 & 19.6 & \cite{Zhang:2012mp} & 0.4783 & 83.8 & 10.2 & \cite{Moresco:2016mzx}\\
0.09 & 69 & 12 & \cite{Jimenez:2003iv} & 0.48 & 97.0 & 62.0 & \cite{Stern:2009ep}\\
0.12 & 68.6 & 26.2 & \cite{Zhang:2012mp} & 0.5929 & 107.0 & 15.5 & \cite{Moresco:2012jh}\\
0.17 & 83.0 & 8.0 & \cite{Simon:2004tf} & 0.6797 & 95.0 & 10.5 & \cite{Moresco:2012jh}\\
0.1791 & 78.0 & 6.2 & \cite{Moresco:2012jh} & 0.75 & 98.8 & 33.6 & \cite{Borghi_2022}\\
0.1993 & 78.0 & 6.9 & \cite{Moresco:2012jh} & 0.7812 & 96.5 & 12.5 & \cite{Moresco:2012jh}\\
0.20 & 72.9 & 29.6 & \cite{Zhang:2012mp} & 0.8754 & 124.5 & 17.4 & \cite{Moresco:2012jh}\\
0.27 & 77.0 & 14.0 & \cite{Simon:2004tf} & 0.88 & 90.0 & 40.0 & \cite{Stern:2009ep}\\
0.28 & 88.8 & 36.6 & \cite{Zhang:2012mp} & 0.9 & 117 & 23 & \cite{Simon:2004tf}\\
0.3519 & 85.5 & 15.7 & \cite{Moresco:2012jh} & 1.037 & 133.5 & 17.6 & \cite{Moresco:2012jh}\\
0.3802 & 83 & 13.5 & \cite{Moresco:2016mzx} & 1.3 & 168.0 & 17.0 & \cite{Simon:2004tf}\\
0.4 & 95 & 17 & \cite{Simon:2004tf} & 1.363 & 160.0 & 33.8 & \cite{Moresco:2015cya}\\
0.4004 & 79.9 & 11.4 & \cite{Moresco:2016mzx} & 1.43 & 177.0 & 18.0 & \cite{Simon:2004tf}\\
0.4247 & 90.4 & 12.8 & \cite{Moresco:2016mzx} & 1.53 & 140.0 & 14.0 & \cite{Simon:2004tf}\\
0.4497 & 96.3 & 14.4 & \cite{Moresco:2016mzx} & 1.75 & 202.0 & 40.0 & \cite{Simon:2004tf}\\
0.47 & 89.0 & 49.6 & \cite{Ratsimbazafy:2017vga} & 1.965 & 186.5 & 50.6 & \cite{Moresco:2015cya}\\
\hline
\end{tabular}

\end{table}
\subsection{Corrected supernova magnitude}

To avoid assuming light curve corrections calibrated within a $\Lambda$CDM framework, we fit the supernova nuisance parameters $(\alpha,\beta,\gamma,M_0)$ simultaneously with the cosmological parameters, following \cite{PhysRevD.111.083508}. 

For each Type Ia supernova the standardized apparent magnitude is constructed using the SALT2 light-curve parameters,
\begin{equation}
m_{B,i}^{\rm corr}=m_{B,i}+\alpha\, x_{1,i}-\beta\, c_i+\gamma\, G_{{\rm host},i},
\label{eq:mbcorr}
\end{equation}
where $m_{B,i}$ is the observed peak magnitude, $x_{1,i}$ and $c_i$ are the stretch and color parameters, and $G_{{\rm host},i}$ accounts for the host-galaxy mass step,
\begin{equation}
G_{{\rm host},i} =
\begin{cases}
+0.5, & \log_{10}(M_{\rm host}/M_\odot) \ge 10, \\
-0.5, & \log_{10}(M_{\rm host}/M_\odot) < 10 .
\end{cases}
\end{equation}

For Hubble-flow supernovae, the theoretical distance modulus is
\begin{equation}
\mu_{\rm th}(z,\theta)=5\log_{10}\!\left[\frac{d_L(z,\theta)}{\rm Mpc}\right]+25 ,
\end{equation}
where $d_L(z,\theta)$ is the luminosity distance for cosmological parameters $\theta$. The residual vector is
\begin{equation}
\Delta\mu_i^{\rm SN}=m_{B,i}^{\rm corr}-M_0-\mu_{\rm th}(z_i,\theta),
\end{equation}
with corresponding $\chi^2$
\begin{equation}
\chi^2_{\rm HF}=(\Delta\mu^{\rm SN})^{T}\mathbf{C}_{\rm SN}^{-1}\Delta\mu^{\rm SN},
\end{equation}
where $\mathbf{C}_{\rm SN}$ is the Pantheon+ covariance matrix restricted to the Hubble-flow sample.

For Cepheid-host calibrator supernovae, the distance modulus is taken from the SH0ES distance ladder,
\begin{equation}
\Delta\mu_i^{\rm Cph}=m_{B,i}^{\rm corr}-M_0-\mu_i^{\rm SH0ES},
\end{equation}
with $\chi^2$
\begin{equation}
\chi^2_{\rm Cph}=(\Delta\mu^{\rm Cph})^{T}\mathbf{C}_{\rm Cph}^{-1}\Delta\mu^{\rm Cph}.
\end{equation}

\subsection{Total likelihood and parameter estimation}
\label{sec:likelihood}

The total supernova contribution is given by
\begin{equation}
\chi^2_{\rm SN}=\chi^2_{\rm HF}+\chi^2_{\rm Cph},
\end{equation}
The full parameter vector sampled in the analysis is
\begin{equation}
\Theta=\{\theta_{\rm cosmo},\alpha,\beta,\gamma,M_0\},
\end{equation}
where $\theta_{\rm cosmo}$ denotes the cosmological parameters and $(\alpha,\beta,\gamma,M_0)$ are the supernova nuisance parameters.

Joint constraints are obtained by combining all datasets,
\begin{equation}
\chi^2_{\rm tot}=
\chi^2_{H(z)}+
\chi^2_{\rm SN}+
\chi^2_{\rm BAO}+
\chi^2_{\rm CMB},
\end{equation}
corresponding to the likelihood
\begin{equation}
\mathcal{L}_{\rm tot}=
\mathcal{L}_{H(z)}\,\mathcal{L}_{\rm SN}\,\mathcal{L}_{\rm BAO}\,\mathcal{L}_{\rm CMB}.
\end{equation}

For the interacting models considered here, the primary cosmological parameter set is
\begin{equation}
\{H_0,\Omega_{b0},\Omega_{\rm dm0},\xi\},
\end{equation}
The radiation density parameter is fixed to its standard value.
We explore the posterior distribution of the model parameters using the \texttt{emcee} Python package, an affine invariant ensemble sampler for Markov Chain Monte Carlo (MCMC) methods~\cite{ForemanMackey2013}.

\section{Background constraints and model comparison}
\label{sec:background_constraints}
\begin{figure*}[ht!]
\centering
\includegraphics[width=0.8\textwidth]{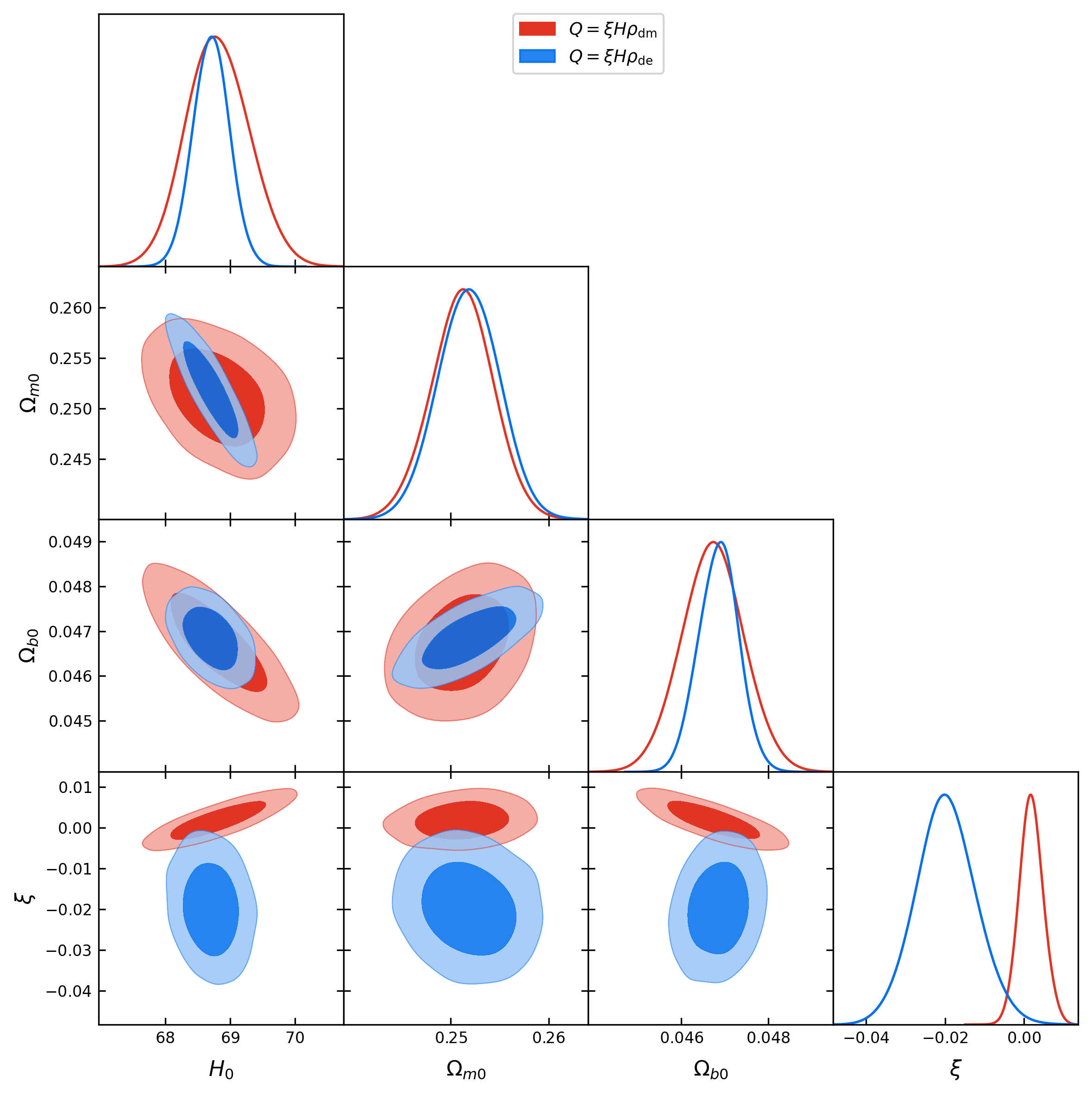}
\caption{Marginalized one and two-dimensional posterior distributions for
$H_0$, $\Omega_{m0}$, $\Omega_{b0}$, and the interaction parameter $\xi$
obtained using background datasets ($H(z)$, Pantheon+ SNe, BAO, and CMB
distance priors). The contours represent the $68\%$ and $95\%$ confidence
regions for the two interaction prescriptions,
$Q=\xi H\rho_{\rm dm}$ and $Q=\xi H\rho_{\rm de}$, used to represent the
diffusion model at the background level.}
\label{fig:corner}
\end{figure*}

In Fig.~\ref{fig:corner}, we present the marginalized one and two dimensional posterior distributions of the cosmological parameters $\{H_0, \Omega_{m0}, \Omega_{b0}, \xi\}$ obtained from a joint analysis of
$H(z)$, BAO DESI DR2, Pantheon+ SNIa, and CMB distance priors.
The red contours correspond to the vacuum proportional interaction
$Q=\xi H\rho_{\mathrm{de}}$, while the blue contours represent the
matter-proportional interaction $Q=\xi H\rho_{\mathrm{dm}}$. For comparison, we also considered the standard $\Lambda$CDM model and the constant equation of state $w$CDM model. A quantitative summary of the parameter constraints and model comparison statistics is presented in Table~\ref{tab:results}.
\begin{table*}[t]
\centering
\caption{Mean values and $1\sigma$ uncertainties for the model parameters
in the interacting dark sector models, $\Lambda$CDM, and $w$CDM obtained
from a joint analysis of $H(z)$, BAO DESI DR2, Pantheon+ SNIa, and CMB distance
priors. The minimum $\chi^2$ values and the corresponding differences
$\Delta\chi^2$ and $\Delta\mathrm{AIC}$ are quoted relative to $\Lambda$CDM.}
\label{tab:results}
\begin{tabular}{lcccc}
\hline\hline
Parameter
& $Q=\xi H\rho_{\rm dm}$
& $Q=\xi H\rho_{\rm de}$
& $\Lambda$CDM
& $\omega$CDM\\
\hline
$H_0$                       & $68.81\pm 0.49 $   & $68.71\pm 0.28$   & $68.56\pm 0.27$ &$69.05\pm 0.51 $ \\
$M_0$                       & $-19.367\pm 0.015 $& $-19.3700\pm 0.0091$& $-19.3747\pm 0.0087 $ &$-19.364\pm 0.013 $   \\
$\alpha$                    & $0.1388\pm 0.0042 $& $0.1388\pm 0.0043$& $0.1389\pm 0.0043$ & $0.1391\pm 0.0042$ \\
$\beta$                     & $2.510\pm 0.073$  & $2.514\pm 0.074$  & $2.513\pm 0.072 $ & $2.519\pm 0.074 $ \\
$\gamma$                    & $0.036\pm 0.010 $  & $0.037\pm 0.010   $  & $0.037\pm 0.010   $ & $0.036\pm 0.010  $ \\
$\Omega_{\rm dm0}$          & $0.2512\pm 0.0032 $& $0.2518\pm 0.0031 $& $0.2509\pm 0.0031  $& $0.2486\pm 0.0038  $ \\
$\Omega_{b0}$               & $0.04672\pm 0.00072$  & $0.04686\pm 0.00046$ & $0.04705\pm 0.00046$ & $0.04635\pm 0.00077 $ \\
$\omega_{\rm de}$ (fixed)   & $-1$ & $-1$ & $-1$ & $-1.025\pm 0.021 $\\
$\xi$                       & $0.0018\pm 0.0031 $ & $-0.0197\pm 0.0076 $  & $0$ & $0$ \\
\hline
$\chi^2_{\min}$             & $1641.3$ & $1634.5$ & $1640.8$ & $1641.7$ \\
$\Delta \chi^2$             & $0.5$    & $-6.3$    & $0$ & $0.9$ \\
$\Delta \mathrm{AIC}$       & $2.5$    & $-4.3$    & $0$ & $2.9$\\
\hline\hline
\end{tabular}
\end{table*}

The standard cosmological parameters $H_0$, $\Omega_{m0}$, and $\Omega_{b0}$ are
tightly constrained and show substantial overlap between the two interaction
models. This demonstrates that background cosmological data robustly determine the global expansion parameters largely independently of the detailed modeling of the dark sector.
In particular, the inferred values of $H_0$ cluster around
$H_0 \simeq 68\,\mathrm{km\,s^{-1}\,Mpc^{-1}}$ for all models considered, with no
significant shifts induced by the inclusion of an interaction or a free dark
energy equation of state.

In comparison, the effective coupling parameter $\xi$ shows a clear dependence on
the assumed interaction prescription. For the matter-proportional interaction
$Q=\xi H\rho_{\mathrm{dm}}$, we obtain
$\xi=0.0018\pm 0.0031$, which is tightly consistent with zero and indicates that
background observations strongly suppress deviations from standard dark matter
conservation. For the vacuum-proportional interaction $Q=\xi H\rho_{\mathrm{de}}$,
the posterior distribution is broader, giving
$\xi=-0.0197\pm 0.0076$, with a mild preference for negative values corresponding
to a decaying vacuum and energy transfer from dark energy to dark matter.
Nevertheless, in both cases the coupling remains consistent with zero within
$2\sigma$, and no statistically significant evidence for a dark sector
interaction is found.

 Among the models considered, the interacting vacuum model with $Q \propto \rho_{\mathrm{de}}$ gives the lowest $\chi^2_{\min}$, improving the fit relative to $\Lambda$CDM by $\Delta\chi^2 = -6.3$. According to the Akaike Information Criterion (AIC), this corresponds to $\Delta \mathrm{AIC} = -4.3$, indicating a moderate statistical preference for this model. In contrast, the $Q \propto \rho_{\mathrm{dm}}$ model and the $w$CDM extension provide fits comparable to $\Lambda$CDM but are mildly disfavored once parameter complexity is taken into account. While the statistical improvement for the vacuum transfer case is non-negligible, it remains modest and does not constitute decisive evidence against $\Lambda$CDM.

Taken together, these results show that unimodular diffusion, interacting vacuum
models with $Q=\xi H\rho_{\mathrm{dm}}$ or $Q=\xi H\rho_{\mathrm{de}}$, $\Lambda$CDM,
and constant $w$ dark energy models with $w\simeq -1$ form a single equivalence
class at the background level. Current background data constrain the expansion
history of the universe with high precision, but are intrinsically insensitive
to the underlying microphysical origin of dark energy or dark sector
interactions. Differences in the inferred values of the coupling parameter $\xi$
reflect the model-dependent definition of the interaction term rather than
physically distinct expansion dynamics.

\section{Linear Perturbations and Growth of Structure}
\label{sec:perturbations}

In the previous section we showed that unimodular diffusion, interacting vacuum models, $\Lambda$CDM, and $w$CDM form an equivalence class at the background level once the effective interaction term is appropriately identified. Although current background observations tightly constrain the expansion history $H(z)$, they do not uniquely determine the microphysical origin of dark energy or dark sector interactions. To break this degeneracy, it is necessary to examine the evolution of cosmological perturbations. The growth of structure provides an independent and sensitive probe of energy exchange in the dark sector, since even small modifications to the conservation equations can alter the clustering dynamics of matter. In this section, we therefore analyze scalar perturbations and derive the corresponding growth equation, with particular emphasis on identifying the conditions under which unimodular diffusion remains perturbatively equivalent to phenomenological interacting vacuum models. We study scalar perturbations around a spatially flat Friedmann-Lemaitre-Robertson-Walker (FLRW) background.
In the Newtonian gauge the perturbed metric is
\begin{equation}
ds^{2}=-(1+2\Psi)dt^{2}+a^{2}(t)(1-2\Phi)\delta_{ij}dx^{i}dx^{j}.
\label{eq:metric}
\end{equation}

The unimodular field equations can be rewritten in the Einstein form
\begin{equation}
G_{\mu\nu}=8\pi G\left(T_{\mu\nu}+T^{(\Lambda)}_{\mu\nu}\right),
\label{eq:einstein_form}
\end{equation}
where the effective vacuum component is defined through an integration function $P(t)$
\begin{equation}
\rho_{\Lambda}^{\rm eff}= \frac{\Lambda}{8 \pi G} +  P(t).
\label{eq:Lambda_eff}
\end{equation}
The bare cosmological constant $\Lambda$ is constant, while $P(t)$ encodes the diffusion process.
Energy-momentum conservation for the total matter plus vacuum is preserved, but the two components exchange energy according to
\begin{equation}
\dot{\rho}_{c}+3H\rho_{c}= - \dot P,\qquad 
\dot{\rho}^{\rm eff}_{\Lambda}= \dot P,
\label{eq:conservation}
\end{equation}

Since our model preserves full diffeomorphism invariance ($\delta\sqrt{-g}\neq0$) and we consider the case diffusion function $P(t)$ purely homogeneous and does not represent an independent dynamical field, its perturbation vanishes identically at linear order. Consequently, 
\begin{equation}
\delta P(t)= \delta \dot P(t)=0,
\label{eq:deltaLambda_zero}
\end{equation} 
and the effective vacuum component remains spatially uniform. 
This is a key point, as a result the perturbed Einstein equations reduce to their standard general relativistic form.  In particular the Poisson equation and the Euler equation for cold dark matter (CDM) read
\begin{align}
\nabla^{2}\Phi &= 4\pi G a^{2}\rho_{m}\delta_{m}, \label{eq:poisson}\\
\dot{\theta}_{c}+H\theta_{c} &= \frac{k^{2}}{a}\Phi, \label{eq:euler}
\end{align}
where $\delta_{m}=(\rho_{c}\delta_{c}+\rho_{b}\delta_{b})/\rho_{m}$ is the total matter density contrast and $\theta_{c}$ is the CDM velocity divergence.  Baryons and radiation follow their usual equations and will be included in the full numerical solution, but for the derivation of the CDM growth equation they can be neglected in a first approximation.

The perturbed part of the conservation equation (\ref{eq:conservation}) for CDM, taking into account the diffusion is,
\begin{equation}
\dot{\delta}_{c}= -\frac{\theta_{c}}{a} +\frac{\dot P}{\rho_{c}}\delta_{c}- \frac{\delta \dot P}{\rho_{c}},
\end{equation}
As discussed above, in our formalism $\delta \dot  P(t)=0$, hence
\begin{equation}
\dot{\delta}_{c}= -\frac{\theta_{c}}{a}+ \Gamma\delta_{c},\qquad 
\Gamma\equiv\frac{\dot P}{\rho_{c}}.
\label{eq:pert_continuity}
\end{equation}
Taking a time derivative of (\ref{eq:pert_continuity}) and using (\ref{eq:euler}) to eliminate $\dot{\theta}_{c}$ gives 
\begin{align}
\ddot{\delta}_{c}&= -\frac{\dot{\theta}_{c}}{a} +\frac{H}{a} \theta_c+  \dot{\Gamma}\delta_{c}+ \Gamma\dot{\delta}_{c}\\
&= -\left(- \frac{H}{a}\theta_{c}+\frac{k^{2}}{a^{2}}\Phi\right) +\frac{H}{a} \theta_c +\dot{\Gamma}\delta_{c}+ \Gamma\dot{\delta}_{c}.
\end{align}
The term $\frac{\theta_{c}}{a}$ can be expressed from (\ref{eq:pert_continuity}) as $\frac{\theta_{c}}{a}= -\dot{\delta}_{c}+ \Gamma\delta_{c}$.  Substituting and using the Poisson equation (\ref{eq:poisson}) to replace $\Phi$, we obtain
\begin{equation}
\ddot{\delta}_{c}+ \bigl(2H-\Gamma\bigr)\dot{\delta}_{c}
-\Bigl[4\pi G\rho_{m}\delta_m +\bigl(2H\Gamma+\dot{\Gamma} \bigr )\delta_{c}\Bigr]=0.
\label{eq:growth_general}
\end{equation}

Equation~(\ref{eq:growth_general}) provides the exact linear evolution equation for the CDM density contrast in the presence of the unimodular diffusion function $P(t)$. Identifying $\dot P(t)=Q(t)$, the resulting growth equation coincides with that of interacting vacuum energy models with equation of state $w=-1$ and interaction $Q \propto \rho_\Lambda$, provided the four-momentum transfer is purely timelike and $\delta\rho_\Lambda=0$ \cite{clemson_2012}.

Importantly, Eq.~(\ref{eq:growth_general}) depends only on the background quantities $H$, $\rho_c$, $\rho_\Lambda$, and the effective interaction rate $\Gamma \equiv Q/\rho_c$, and is therefore insensitive to the microscopic origin of the energy exchange. Any theory yielding the same expansion history $H(z)$ and interaction function $\Gamma(z)$ predicts the same linear evolution of $\delta_c(z)$.

To connect with observations, we relate the growth equation to the observable $f\sigma_{8}$. Defining $f\equiv d\ln\delta_{c}/d\ln a$ and using $d/d\ln a = H^{-1}d/dt$, Eq.~(\ref{eq:growth_general}) can be rewritten as
\begin{equation}
\frac{df}{d\ln a}+ f^{2}+ \left(2+\frac{\dot{H}}{H^{2}}-\frac{\Gamma}{H}\right)f
= \frac{3}{2}\Omega_m+\frac{2\Gamma}{H}+\frac{\dot{\Gamma}}{H^{2}},
\label{eq:f_ode}
\end{equation}

Because Eqs.~(\ref{eq:f_ode})) coincide for unimodular diffusion and interacting vacuum energy models with homogeneous transfer at linear order, constraints derived from $f\sigma_{8}$ measurements apply equally to both frameworks. Within this restricted subclass, growth data constrain only the effective transfer rate $\Gamma(z)$ and do not distinguish between the geometric diffusion and its phenomenological interacting vacuum counterpart. Hence, under the assumptions of homogeneous $P(t)$ and absence of momentum exchange, unimodular diffusion and interacting vacuum energy belong to the same dynamical equivalence class at the level of linear scalar perturbations.

To constrain the modified growth equation with observations, we use redshift-space distortion (RSD) measurements of the quantity $f\sigma_8(z)$, which directly probe the linear growth rate of matter perturbations. We adopt a compilation of independent measurements spanning the redshift range $z \leq 2$ from Ref.~\cite{RSD_2018}, summarized in Table~\ref{tab:rsd_data}. The RSD measurements used in this work are reported assuming a fiducial $\Lambda$CDM cosmology. Since the interacting models considered here produce expansion histories that do not deviate significantly from the fiducial model, the Alcock-Paczynski rescaling is expected to introduce only subdominant corrections compared to current observational uncertainties. For simplicity, we therefore neglect this effect in the present analysis. The RSD data are included in the joint likelihood in order to assess how growth measurements constrain the interaction parameter $\xi$.
\begin{table}[t]
\centering
\caption{Compilation of redshift-space distortion (RSD) measurements of $f\sigma_8(z)$ used in this work. The quoted uncertainties correspond to $1\sigma$ errors. Data are taken from Ref.~\cite{RSD_2018}.}
\begin{tabular}{ccc}
\hline
Dataset & $z$ & $f\sigma_8(z)$  \\
\hline
 6dFGS+SnIa & 0.02  & $0.428  \pm 0.0465$ \\
 6dFGS & 0.067 & $0.423  \pm 0.055$    \\
 SDSS-MGS & 0.15  & $0.490  \pm 0.145$    \\
2dFGRS & 0.17  & $0.510  \pm 0.060$    \\
BOSS-LOWZ & 0.32  & $0.384  \pm 0.095$    \\
SDSS-LRG-200 & 0.37  & $0.4602 \pm 0.0378$   \\
BOSS DR12 & 0.40  & $0.473  \pm 0.086$    \\
 WiggleZ & 0.44  & $0.413  \pm 0.080$    \\
 SDSS-BOSS & 0.50  & $0.427  \pm 0.043$    \\
BOSS CMASS & 0.57  & $0.426  \pm 0.029$    \\
 WiggleZ & 0.60  & $0.390  \pm 0.063$    \\
WiggleZ & 0.73  & $0.437  \pm 0.072$    \\
VVDS & 0.77  & $0.490  \pm 0.180$    \\
Vipers & 0.80  & $0.470  \pm 0.080$    \\
 Vipers PDR-2 & 0.86  & $0.400  \pm 0.110$    \\
Vipers v7 & 1.05  & $0.280  \pm 0.080$   \\
FastSound & 1.40  & $0.482  \pm 0.116$    \\
SDSS-IV & 1.52  & $0.396  \pm 0.079$    \\
 SDSS-IV& 1.944 & $0.364  \pm 0.106$    \\
\hline
\end{tabular}
\label{tab:rsd_data}
\end{table}

The interaction considered here does not modify the gravitational sector and involves purely timelike energy transfer, with homogeneous vacuum perturbations ($\delta P = 0$). Under these assumptions, the linear growth equation for cold dark matter remains scale-independent on subhorizon scales. So, the matter power spectrum can be written as~\cite{Dodelson2003},

\begin{equation}
P_m(k,z)=D^2(z)P_m(k,0),
\end{equation}
where growth factor $D(z)$ is obtained from the definition 
$f = d\ln D/d\ln a$, which gives
\begin{equation}
D(a) = \exp\!\left( \int_{a_i}^{a} f(a')\, d\ln a' \right).
\end{equation}
The normalization is chosen such that $D(a=1)=1$.
The present-day clustering amplitude is parameterized through
\begin{equation}
S_8 = \sigma_{8,0}\sqrt{\frac{\Omega_{m0}}{0.3}},
\end{equation}
with $S_8$ treated as a free parameter. The observable quantities are therefore computed as
\begin{equation}
\sigma_8(z)=\sigma_{8,0}D(z),
\qquad
f\sigma_8(z)=f(z)\sigma_{8,0}D(z),
\end{equation}
where $\sigma_{8,0}=S_8/\sqrt{\Omega_{m0}/0.3}$. 
The background evolution is constrained using CMB distance priors, while the redshift dependence of clustering follows directly from the scale-independent growth factor.

\begin{figure*}[ht]
\centering

\begin{minipage}{0.48\textwidth}
\centering
\includegraphics[width=\linewidth]{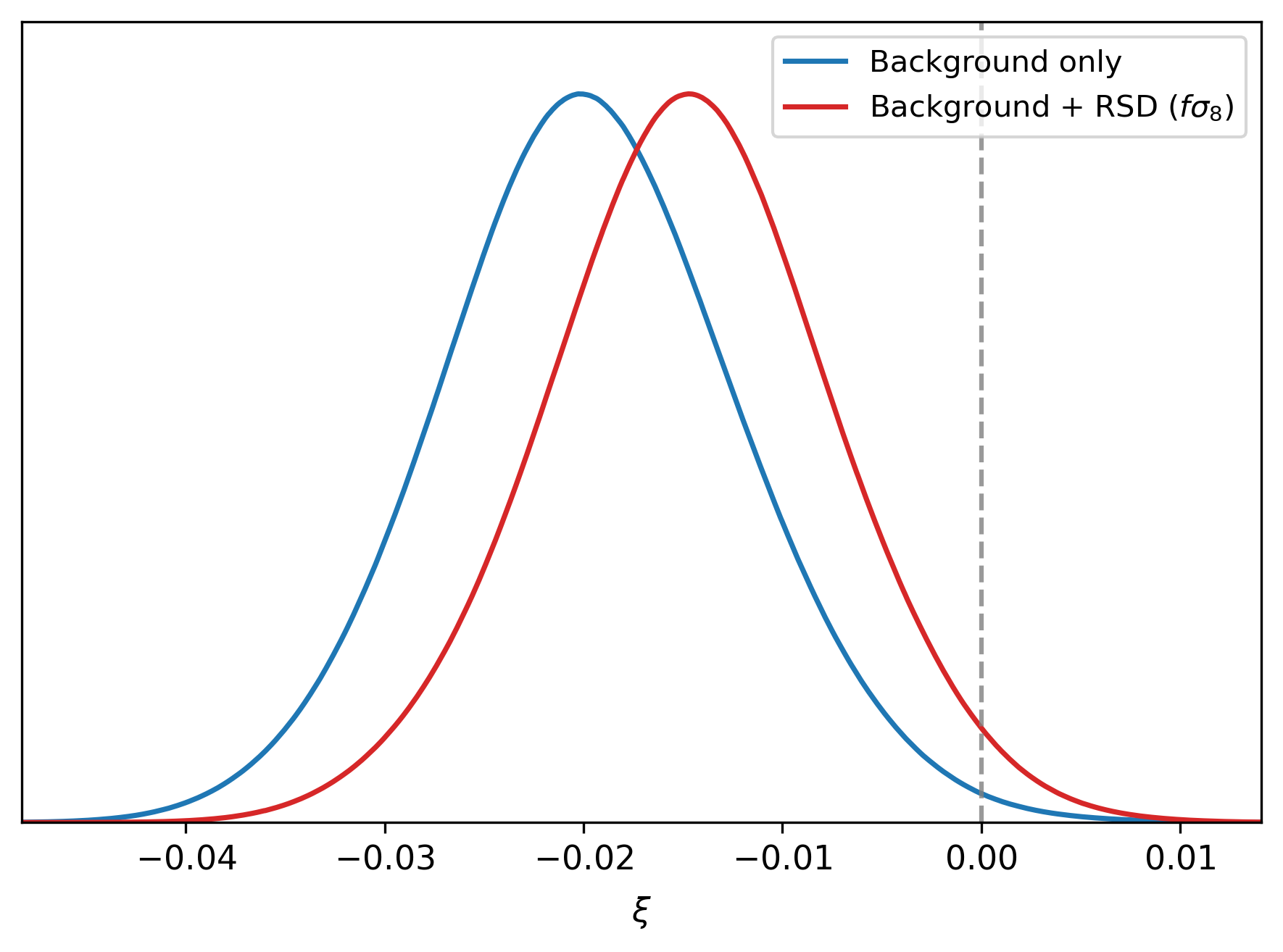}
\end{minipage}
\hfill
\begin{minipage}{0.48\textwidth}
\centering
\includegraphics[width=\linewidth]{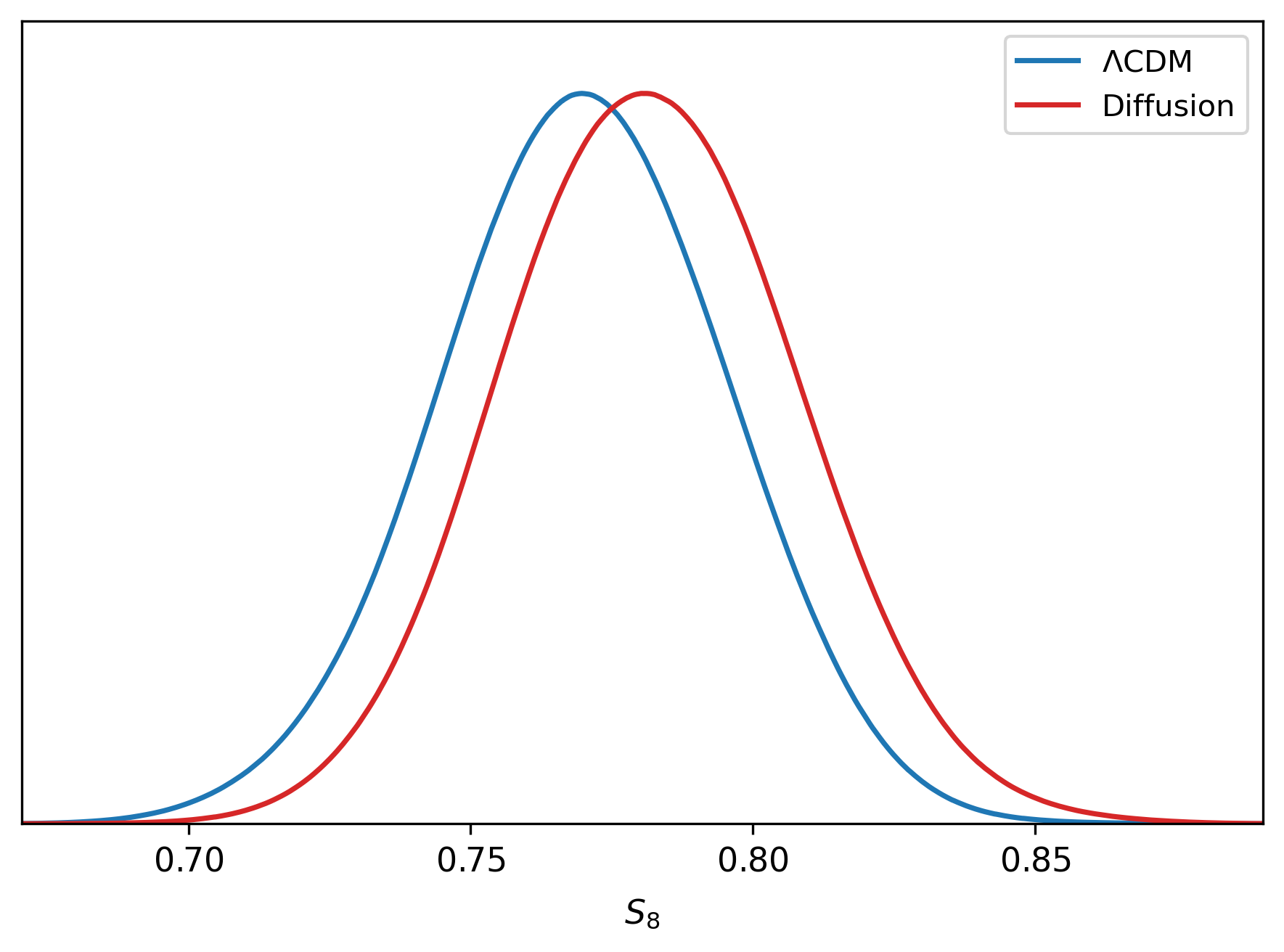}
\end{minipage}
\caption{
Left: One dimensional posterior distributions of the interaction parameter $\xi$ 
for the diffusion model $Q=\xi H \rho_\Lambda$ using background-only data and the combined background+RSD analysis. 
Right: Corresponding constraints on $S_8$ for $\Lambda$CDM and the diffusion model 
using the full dataset. Growth measurements mildly reduce the statistical 
preference for nonzero $\xi$ while inducing only a modest shift in $S_8$.
}
\label{fig:xi_s8_comparison}
\end{figure*}

The impact of including growth measurements on the parameter $\xi$ is illustrated in Fig.~\ref{fig:xi_s8_comparison}. 
For the diffusion model, background-only data yield $\xi = -0.0197 \pm 0.0076$, corresponding to a $\sim 2.6\sigma$ indication of a small negative coupling. 
When $f\sigma_8$ measurements are included, the constraint shifts to $\xi = -0.0146 \pm 0.0075$, reducing the statistical significance to approximately $2\sigma$. 
Correspondingly, the improvement in goodness-of-fit relative to $\Lambda$CDM seen in the background-only analysis ($\Delta\chi^2 = -6.3$) is removed once growth data are included, yielding $\Delta\chi^2 = +1.37$ for the full dataset. 
Thus, the preference for interaction present in the background-only analysis is not maintained once large-scale structure measurements are taken into account.

For the combined dataset considered here, $\Lambda$CDM gives 
$S_8 = 0.77 \pm 0.025$, which is lower than the Planck $\Lambda$CDM value 
$S_8 \approx 0.83$. The diffusion model shifts this value slightly upward to 
$S_8 = 0.782 \pm 0.026$, although it remains below the Planck result. 
This difference mainly reflects the absence of the full CMB power spectrum 
likelihood in our analysis, rather than a strong model-dependent effect. 

\begin{figure*}[htb]
\centering
\includegraphics[width=0.8\textwidth]{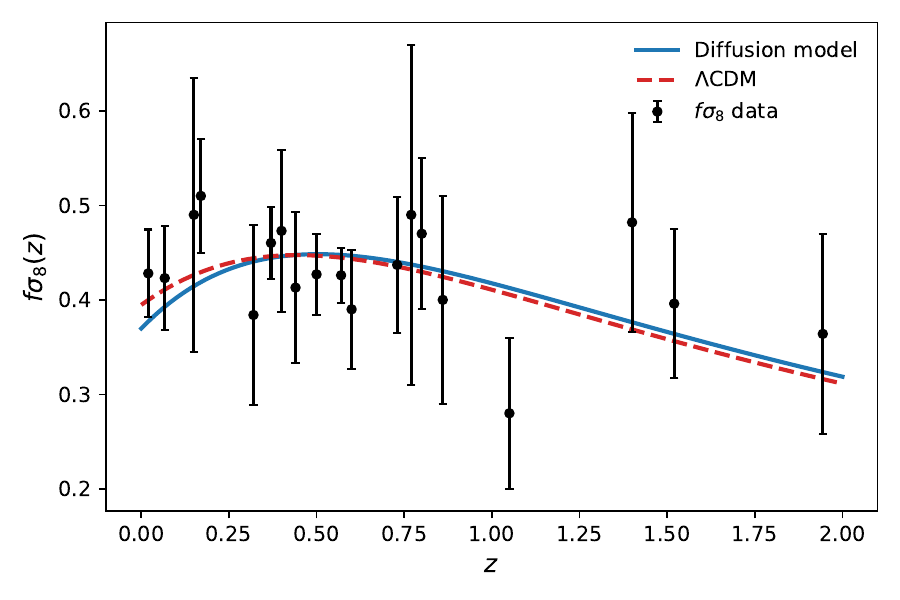}
\caption{
Evolution of the growth-rate observable $f\sigma_8(z)$.
The blue solid curve shows the best-fit prediction of the diffusion model,
while the red dashed curve corresponds to the $\Lambda$CDM prediction obtained
using the same dataset combination.
Black points with error bars represent redshift-space distortion (RSD)
measurements of $f\sigma_8$.
The two models produce nearly identical growth histories within current
observational uncertainties, indicating that present large-scale structure
data do not strongly distinguish between diffusion-driven vacuum dynamics
and the standard $\Lambda$CDM model.
}
\label{fig_fs8}
\end{figure*}

The corresponding predictions for the growth observable $f\sigma_8(z)$ are 
shown in Fig.~\ref{fig_fs8}, where both the diffusion model and the 
$\Lambda$CDM model provide nearly identical growth histories within the 
current observational uncertainties. The diffusion model produces a 
slightly higher clustering amplitude, consistent with the modest upward 
shift in $S_8$, but the difference remains well within the error bars of 
the present redshift-space distortion measurements.
Taken together, these results indicate that although the perturbation 
analysis restricts the background equivalence class to homogeneous 
interacting vacuum models, current growth data do not provide 
statistically significant evidence for a nonzero interaction. 
Consequently, the diffusion model remains observationally consistent with 
$\Lambda$CDM, and the degeneracy between the two descriptions is only 
weakly lifted by present large-scale structure measurements.

It is important to emphasize that the perturbative equivalence established above
does not extend to arbitrary interacting dark sector models, but only to a
restricted subclass. In this work we adopt the fully diffeomorphism-invariant
formulation of unimodular gravity, for which $\delta\sqrt{-g}\neq 0$. In this
formulation the gravitational sector and gauge structure coincide with those of
general relativity, allowing the use of standard Newtonian or synchronous gauge
perturbation theory. Constrained unimodular formulations with
$\delta\sqrt{-g}=0$ possess reduced gauge freedom and lead to modified
perturbation equations, which lie beyond the scope of the present analysis.

In the unimodular diffusion framework the integration function $P(t)$ is purely
homogeneous by construction. Consequently $\delta P = 0$ and, since $Q=\dot P$,
one also has $\delta Q = 0$. The perturbed CDM continuity equation therefore
depends only on the background interaction rate
$\Gamma = Q/\rho_c$ and contains no intrinsic perturbation of the transfer
term.

In phenomenological interacting dark energy models, however, the perturbation
$\delta Q$ depends explicitly on the chosen coupling prescription. For example,
if $Q \propto \rho_c$ one typically obtains
\begin{equation}
\delta Q \propto \delta\rho_c ,
\end{equation}
which directly modifies the matter growth equation at linear order. This
statement assumes the commonly adopted choice of vanishing momentum transfer in
the dark matter rest frame, $Q^\mu = Q u^\mu_{\mathrm{dm}}$, such that the
energy exchange is purely timelike and aligned with the dark matter
four-velocity.

For interacting vacuum models with $Q \propto \rho_\Lambda$,
$w=-1$, and vanishing vacuum perturbations $\delta\rho_\Lambda = 0$, the
perturbed transfer term becomes
\begin{equation}
\delta Q = \xi \rho_\Lambda \delta H .
\end{equation}
On subhorizon scales metric perturbations are suppressed relative to matter
density fluctuations, leaving $\delta H$ and therefore $\delta Q$
subdominant. In this regime the resulting growth equation coincides with that
obtained in the unimodular diffusion framework.

Therefore the perturbative equivalence with unimodular diffusion holds only for
interacting vacuum energy models in which the transfer term remains effectively
homogeneous at linear order in the subhorizon limit. More general interacting
models, where $\delta Q$ inherits clustering from $\delta\rho_c$ or additional
metric perturbations, lead to distinguishable signatures in the growth rate and large-scale structure observables, providing a potential observational test of the diffusion framework. Relaxing any of these assumptions would modify the perturbation equations and break the degeneracy between unimodular
diffusion and interacting vacuum energy. Within the framework considered here,
however, these conditions are satisfied by construction.

\section{Conclusion}
\label{sec:con}
In this work we clarified the correspondence between unimodular diffusion cosmology and interacting dark sector models. By identifying the diffusion rate with an effective energy transfer term, $Q = -\dot P$, we showed that unimodular diffusion maps onto interacting vacuum models with $w_{\rm de} = -1$ at the background level. This mapping implies that the diffusion framework does not generate a distinct expansion history but instead provides a geometric realization of the interacting vacuum class. At the background level, both the vacuum‑coupled ($Q = \xi H \rho_{\rm de}$) and matter‑coupled ($Q = \xi H \rho_{\rm dm}$) parameterizations fit the data comparably. However, at linear perturbation level only the vacuum‑coupled interaction remains consistent with the diffusion framework.

Using cosmic chronometer $H(z)$ data, Pantheon+ supernovae, DESI BAO, and CMB distance priors, we obtained $\xi = -0.0197 \pm 0.0076$ for the vacuum‑coupled case. Including redshift‑space distortion measurements ($f\sigma_8$) shifts the constraint to $\xi = -0.0147 \pm 0.0075$, which remains consistent with $\Lambda$CDM ($\xi = 0$) at the $2\sigma$ level. The corresponding clustering amplitudes, $S_8 = 0.77 \pm 0.025$ for $\Lambda$CDM and $S_8 = 0.782 \pm 0.026$ for the diffusion model (with the same datasets), indicate a modest impact on late-time structure growth.

At the perturbative level, the defining condition of the unimodular diffusion framework is $\delta P = 0$ (equivalently $\delta Q = 0$), which reflects the homogeneity of the diffusion function and the absence of momentum transfer. Under this condition, together with full diffeomorphism invariance, purely timelike energy transfer, and the sub‑horizon linear regime, the CDM growth equation in unimodular diffusion coincides exactly with that of interacting vacuum models sharing the same $Q(t)$. Consequently, linear observables, $f$, $D(z)$, and $f\sigma_8(z)$, depend only on the effective transfer rate $\Gamma = Q/\rho_c$ and the background expansion $H(z)$, making them insensitive to the microscopic origin of the interaction. Within the linear regime, unimodular diffusion and homogeneous interacting vacuum models thus belong to the same dynamical equivalence class. The matter‑coupled interaction, on the other hand, does not satisfy $\delta Q = 0$ and therefore does not represent a perturbatively consistent class of the diffusion model.

Our results indicate that current cosmological observations based on the homogeneous expansion history and linear perturbations constrain only the effective energy exchange in the dark sector. They do not differentiate between its geometric realization in unimodular diffusion and its phenomenological description in interacting vacuum models. Although linear perturbations restrict the background equivalence class to homogeneous vacuum transfer models, present growth measurements do not break the degeneracy with $\Lambda$CDM.

Future efforts to differentiate these frameworks will likely require probes sensitive to momentum transfer, non‑linear structure formation, or deviations from standard gravitational dynamics beyond the linear regime. Hence, unimodular diffusion stands as a consistent and perturbatively stable extension of standard cosmology that remains compatible with current data while highlighting the theoretical degeneracies inherent in linear cosmological observations.

\section*{Acknowledgements}
GK acknowledge Vellore Institute of Technology for the financial support through its Seed Grant (No.SG20230035), year 2023. 

%------------------------
%%%%%%%%%%%%%%%%%%%%%%%%%%%
% \section*{Acknowledgements}
%\end{acknowledgements}
%%%%%%%%%%%%%%%%%%%%%%%%%%%

 \bibliographystyle{JHEP}
  \bibliography{INTDEDM}

@article{RSD_2018,
  title = {Evolution of the $f{\ensuremath{\sigma}}_{8}$ tension with the $\text{Planck}15/\mathrm{\ensuremath{\Lambda}}\mathrm{CDM}$ determination and implications for modified gravity theories},
  author = {Kazantzidis, Lavrentios and Perivolaropoulos, Leandros},
  journal = {Phys. Rev. D},
  volume = {97},
  issue = {10},
  pages = {103503},
  numpages = {16},
  year = {2018},
  month = {May},
  publisher = {American Physical Society},
  doi = {10.1103/PhysRevD.97.103503},
  url = {https://link.aps.org/doi/10.1103/PhysRevD.97.103503}
}

@article{Luca_2000,
  title = {Coupled quintessence},
  author = {Amendola, Luca},
  journal = {Phys. Rev. D},
  volume = {62},
  issue = {4},
  pages = {043511},
  numpages = {10},
  year = {2000},
  month = {Jul},
  publisher = {American Physical Society},
  doi = {10.1103/PhysRevD.62.043511},
  url = {https://link.aps.org/doi/10.1103/PhysRevD.62.043511}
}

@article{PhysRevD.74.023501,
  title = {Interacting models of soft coincidence},
  author = {del Campo, Sergio and Herrera, Ram\'on and Olivares, Germ\'an and Pav\'on, Diego},
  journal = {Phys. Rev. D},
  volume = {74},
  issue = {2},
  pages = {023501},
  numpages = {9},
  year = {2006},
  month = {Jul},
  publisher = {American Physical Society},
  doi = {10.1103/PhysRevD.74.023501},
  url = {https://link.aps.org/doi/10.1103/PhysRevD.74.023501}
}

@article{ZIMDAHL2001,
title = {Interacting quintessence},
journal = {Physics Letters B},
volume = {521},
number = {3},
pages = {133-138},
year = {2001},
issn = {0370-2693},
doi = {https://doi.org/10.1016/S0370-2693(01)01174-1},
url = {https://www.sciencedirect.com/science/article/pii/S0370269301011741},
author = {Winfried Zimdahl and Diego Pavón and Luis P. Chimento}
}

@article{Perlmutter,
    author = "Perlmutter, S. and others",
    collaboration = "Supernova Cosmology Project",
    title = "{Measurements of $\Omega$ and $\Lambda$ from 42 High Redshift Supernovae}",
    eprint = "astro-ph/9812133",
    archivePrefix = "arXiv",
    reportNumber = "LBNL-41801, LBL-41801",
    doi = "10.1086/307221",
    journal = "Astrophys. J.",
    volume = "517",
    pages = "565--586",
    year = "1999"
}

@article{Riess,
    author = "Riess, Adam G. and others",
    collaboration = "Supernova Search Team",
    title = "{Observational evidence from supernovae for an accelerating universe and a cosmological constant}",
    eprint = "astro-ph/9805201",
    archivePrefix = "arXiv",
    doi = "10.1086/300499",
    journal = "Astron. J.",
    volume = "116",
    pages = "1009--1038",
    year = "1998"
}

@article{PhysRevD.108.043524,
  title = {Cosmological constraints on unimodular gravity models with diffusion},
  author = {Landau, Susana J. and Benetti, Micol and Perez, Alejandro and Sudarsky, Daniel},
  journal = {Phys. Rev. D},
  volume = {108},
  issue = {4},
  pages = {043524},
  numpages = {16},
  year = {2023},
  month = {Aug},
  publisher = {American Physical Society},
  doi = {10.1103/PhysRevD.108.043524},
  url = {https://link.aps.org/doi/10.1103/PhysRevD.108.043524}
}

@article{Planck:2018vyg,
    author = "Aghanim, N. and others",
    collaboration = "Planck",
    title = "{Planck 2018 results. VI. Cosmological parameters}",
    eprint = "1807.06209",
    archivePrefix = "arXiv",
    primaryClass = "astro-ph.CO",
    doi = "10.1051/0004-6361/201833910",
    journal = "Astron. Astrophys.",
    volume = "641",
    pages = "A6",
    year = "2020",
    note = "[Erratum: Astron.Astrophys. 652, C4 (2021)]"
}

@article{clemson_2012,
  title = {Interacting dark energy: Constraints and degeneracies},
  author = {Clemson, Timothy and Koyama, Kazuya and Zhao, Gong-Bo and Maartens, Roy and V\"aliviita, Jussi},
  journal = {Phys. Rev. D},
  volume = {85},
  issue = {4},
  pages = {043007},
  numpages = {12},
  year = {2012},
  month = {Feb},
  publisher = {American Physical Society},
  doi = {10.1103/PhysRevD.85.043007},
  url = {https://link.aps.org/doi/10.1103/PhysRevD.85.043007}
}

@article{DESI:2024mwx,
    author = "Adame, A. G. and others",
    collaboration = "DESI",
    title = "{DESI 2024 VI: cosmological constraints from the measurements of baryon acoustic oscillations}",
    eprint = "2404.03002",
    archivePrefix = "arXiv",
    primaryClass = "astro-ph.CO",
    reportNumber = "FERMILAB-PUB-24-0154-PPD",
    doi = "10.1088/1475-7516/2025/02/021",
    journal = "JCAP",
    volume = "02",
    pages = "021",
    year = "2025"
}

@article{DESI:2025zgx,
    author = "Abdul Karim, M. and others",
    collaboration = "DESI",
    title = "{DESI DR2 Results II: Measurements of Baryon Acoustic Oscillations and Cosmological Constraints}",
    eprint = "2503.14738",
    archivePrefix = "arXiv",
    primaryClass = "astro-ph.CO",
    reportNumber = "FERMILAB-PUB-25-0169-PPD",
    month = "3",
    year = "2025"
}

@article{Riess:2020fzl,
    author = "Riess, Adam G. and Casertano, Stefano and Yuan, Wenlong and Bowers, J. Bradley and Macri, Lucas and Zinn, Joel C. and Scolnic, Dan",
    title = "{Cosmic Distances Calibrated to 1\% Precision with Gaia EDR3 Parallaxes and Hubble Space Telescope Photometry of 75 Milky Way Cepheids Confirm Tension with $\Lambda$CDM}",
    eprint = "2012.08534",
    archivePrefix = "arXiv",
    primaryClass = "astro-ph.CO",
    doi = "10.3847/2041-8213/abdbaf",
    journal = "Astrophys. J. Lett.",
    volume = "908",
    number = "1",
    pages = "L6",
    year = "2021"
}

@article{H0LiCOW:2018tyj,
    author = "Birrer, S. and others",
    collaboration = "H0LiCOW",
    title = "{H0LiCOW - IX. Cosmographic analysis of the doubly imaged quasar SDSS 1206+4332 and a new measurement of the Hubble constant}",
    eprint = "1809.01274",
    archivePrefix = "arXiv",
    primaryClass = "astro-ph.CO",
    doi = "10.1093/mnras/stz200",
    journal = "Mon. Not. Roy. Astron. Soc.",
    volume = "484",
    pages = "4726",
    year = "2019"
}

@article{Guo:2021rrz,
    author = "Guo, Rui-Yun and Feng, Lu and Yao, Tian-Ying and Chen, Xing-Yu",
    title = "{Exploration of interacting dynamical dark energy model with interaction term including the equation-of-state parameter: alleviation of the H$_{0}$ tension}",
    eprint = "2110.02536",
    archivePrefix = "arXiv",
    primaryClass = "gr-qc",
    doi = "10.1088/1475-7516/2021/12/036",
    journal = "JCAP",
    volume = "12",
    number = "12",
    pages = "036",
    year = "2021"
}

@article{Murgia:2016ccp,
    author = "Murgia, Riccardo and Gariazzo, Stefano and Fornengo, Nicolao",
    title = "{Constraints on the Coupling between Dark Energy and Dark Matter from CMB data}",
    eprint = "1602.01765",
    archivePrefix = "arXiv",
    primaryClass = "astro-ph.CO",
    doi = "10.1088/1475-7516/2016/04/014",
    journal = "JCAP",
    volume = "04",
    pages = "014",
    year = "2016"
}

@article{Hu:2006ar,
    author = "Hu, Bo and Ling, Yi",
    title = "{Interacting dark energy, holographic principle and coincidence problem}",
    eprint = "hep-th/0601093",
    archivePrefix = "arXiv",
    doi = "10.1103/PhysRevD.73.123510",
    journal = "Phys. Rev. D",
    volume = "73",
    pages = "123510",
    year = "2006"
}

@article{Sadjadi:2006qp,
    author = "Sadjadi, H. Mohseni and Alimohammadi, M.",
    title = "{Cosmological coincidence problem in interactive dark energy models}",
    eprint = "gr-qc/0610080",
    archivePrefix = "arXiv",
    doi = "10.1103/PhysRevD.74.103007",
    journal = "Phys. Rev. D",
    volume = "74",
    pages = "103007",
    year = "2006"
}

@article{Berger:2006db,
    author = "Berger, Micheal S. and Shojaei, Hamed",
    title = "{Interacting dark energy and the cosmic coincidence problem}",
    eprint = "gr-qc/0601086",
    archivePrefix = "arXiv",
    reportNumber = "IUHET-490",
    doi = "10.1103/PhysRevD.73.083528",
    journal = "Phys. Rev. D",
    volume = "73",
    pages = "083528",
    year = "2006"
}

@article{Li:2024qso,
    author = "Li, Tian-Nuo and Wu, Peng-Ju and Du, Guo-Hong and Jin, Shang-Jie and Li, Hai-Li and Zhang, Jing-Fei and Zhang, Xin",
    title = "{Constraints on Interacting Dark Energy Models from the DESI Baryon Acoustic Oscillation and DES Supernovae Data}",
    eprint = "2407.14934",
    archivePrefix = "arXiv",
    primaryClass = "astro-ph.CO",
    doi = "10.3847/1538-4357/ad87f0",
    journal = "Astrophys. J.",
    volume = "976",
    number = "1",
    pages = "1",
    year = "2024"
}

@article{Pan:2025qwy,
    author = "Pan, Supriya and Paul, Sivasish and Saridakis, Emmanuel N. and Yang, Weiqiang",
    title = "{Interacting dark energy after DESI DR2: a challenge for $\Lambda$CDM paradigm?}",
    eprint = "2504.00994",
    archivePrefix = "arXiv",
    primaryClass = "astro-ph.CO",
    month = "4",
    year = "2025"
}

@article{Silva:2025hxw,
    author = "Silva, Emanuelly and Sabogal, Miguel A. and Souza, Mateus S. and Nunes, Rafael C. and Di Valentino, Eleonora and Kumar, Suresh",
    title = "{New Constraints on Interacting Dark Energy from DESI DR2 BAO Observations}",
    eprint = "2503.23225",
    archivePrefix = "arXiv",
    primaryClass = "astro-ph.CO",
    month = "3",
    year = "2025"
}

@article{Pandey:2019plg,
    author = "Pandey, Kanhaiya L. and Karwal, Tanvi and Das, Subinoy",
    title = "{Alleviating the $H_0$ and $\sigma_8$ anomalies with a decaying dark matter model}",
    eprint = "1902.10636",
    archivePrefix = "arXiv",
    primaryClass = "astro-ph.CO",
    doi = "10.1088/1475-7516/2020/07/026",
    journal = "JCAP",
    volume = "07",
    pages = "026",
    year = "2020"
}

@book{Dodelson2003,
  author    = {Scott Dodelson},
  title     = {Modern Cosmology},
  publisher = {Elsevier Science Publishing Co Inc},
  year      = {2003},
  isbn      = {9780122191411}
}

@article{PhysRevD.111.083508,
  title = {Anisotropic universe with anisotropic dark energy},
  author = {Verma, Anshul and Aluri, Pavan K. and Mota, David F.},
  journal = {Phys. Rev. D},
  volume = {111},
  issue = {8},
  pages = {083508},
  numpages = {14},
  year = {2025},
  month = {Apr},
  publisher = {American Physical Society},
  doi = {10.1103/PhysRevD.111.083508},
  url = {https://link.aps.org/doi/10.1103/PhysRevD.111.083508}
}

@article{Scolnic_2022,
doi = {10.3847/1538-4357/ac8b7a},
url = {https://doi.org/10.3847/1538-4357/ac8b7a},
year = {2022},
month = {oct},
publisher = {The American Astronomical Society},
volume = {938},
number = {2},
pages = {113},
author = {Scolnic, Dan and Brout, Dillon and Carr, Anthony and Riess, Adam G. and Davis, Tamara M. and Dwomoh, Arianna and Jones, David O. and Ali, Noor and Charvu, Pranav and Chen, Rebecca and Peterson, Erik R. and Popovic, Brodie and Rose, Benjamin M. and Wood, Charlotte M. and Brown, Peter J. and Chambers, Ken and Coulter, David A. and Dettman, Kyle G. and Dimitriadis, Georgios and Filippenko, Alexei V. and Foley, Ryan J. and Jha, Saurabh W. and Kilpatrick, Charles D. and Kirshner, Robert P. and Pan, Yen-Chen and Rest, Armin and Rojas-Bravo, Cesar and Siebert, Matthew R. and Stahl, Benjamin E. and Zheng, WeiKang},
title = {The Pantheon+ Analysis: The Full Data Set and Light-curve Release},
journal = {The Astrophysical Journal}
}

@article{Jimenez_2002,
doi = {10.1086/340549},
url = {https://doi.org/10.1086/340549},
year = {2002},
month = {jul},
publisher = {},
volume = {573},
number = {1},
pages = {37},
author = {Jimenez, Raul and Loeb, Abraham},
title = {Constraining Cosmological Parameters Based on Relative Galaxy Ages},
journal = {The Astrophysical Journal}
}

@article{desi_dr2,
  title = {DESI DR2 results. II. Measurements of baryon acoustic oscillations and cosmological constraints},
  author = {Abdul Karim, M. and Aguilar, J. and Ahlen, S. and Alam, S. and Allen, L. and Prieto, C. Allende and Alves, O. and Anand, A. and Andrade, U. and Armengaud, E. and Aviles, A. and Bailey, S. and Baltay, C. and Bansal, P. and Bault, A. and Behera, J. and BenZvi, S. and Bianchi, D. and Blake, C. and Brieden, S. and Brodzeller, A. and Brooks, D. and Buckley-Geer, E. and Burtin, E. and Calderon, R. and Canning, R. and Rosell, A. Carnero and Carrilho, P. and Casas, L. and Castander, F. J. and Charles, M. and Chaussidon, E. and Chaves-Montero, J. and Chebat, D. and Chen, X. and Claybaugh, T. and Cole, S. and Cooper, A. P. and Cuceu, A. and Dawson, K. S. and de la Macorra, A. and de Mattia, A. and Deiosso, N. and Della Costa, J. and Demina, R. and Dey, A. and Dey, B. and Ding, Z. and Doel, P. and Edelstein, J. and Eisenstein, D. J. and Elbers, W. and Fagrelius, P. and Fanning, K. and Fern\'andez-Garc\'{\i}a, E. and Ferraro, S. and Font-Ribera, A. and Forero-Romero, J. E. and Frenk, C. S. and Garcia-Quintero, C. and Garrison, L. H. and Gazta\~naga, E. and Gil-Mar\'{\i}n, H. and Gontcho, S. Gontcho A. and Gonzalez, D. and Gonzalez-Morales, A. X. and Gordon, C. and Green, D. and Gutierrez, G. and Guy, J. and Hadzhiyska, B. and Hahn, C. and He, S. and Herbold, M. and Herrera-Alcantar, H. K. and Ho, M.-F. and Honscheid, K. and Howlett, C. and Huterer, D. and Ishak, M. and Juneau, S. and Kamble, N. V. and Kara\ifmmode \mbox{\c{c}}\else \c{c}\fi{}ayl��, N. G. and Kehoe, R. and Kent, S. and Kim, A. G. and Kirkby, D. and Kisner, T. and Koposov, S. E. and Kremin, A. and Krolewski, A. and Lahav, O. and Lamman, C. and Landriau, M. and Lang, D. and Lasker, J. and Le Goff, J. M. and Le Guillou, L. and Leauthaud, A. and Levi, M. E. and Li, Q. and Li, T. S. and Lodha, K. and Lokken, M. and Lozano-Rodr\'{\i}guez, F. and Magneville, C. and Manera, M. and Martini, P. and Matthewson, W. L. and Meisner, A. and Mena-Fern\'andez, J. and Menegas, A. and Mergulh\~ao, T. and Miquel, R. and Moustakas, J. and Mu\~noz-Guti\'errez, A. and Mu\~noz-Santos, D. and Myers, A. D. and Nadathur, S. and Naidoo, K. and Napolitano, L. and Newman, J. A. and Niz, G. and Noriega, H. E. and Paillas, E. and Palanque-Delabrouille, N. and Pan, J. and Peacock, J. A. and Ibanez, M. P. and Percival, W. J. and P\'erez-Fern\'andez, A. and P\'erez-R\`afols, I. and Pieri, M. M. and Poppett, C. and Prada, F. and Rabinowitz, D. and Raichoor, A. and Ram\'{\i}rez-P\'erez, C. and Rashkovetskyi, M. and Ravoux, C. and Rich, J. and Rocher, A. and Rockosi, C. and Rohlf, J. and Rom\'an-Herrera, J. O. and Ross, A. J. and Rossi, G. and Ruggeri, R. and Ruhlmann-Kleider, V. and Samushia, L. and Sanchez, E. and Sanders, N. and Schlegel, D. and Schubnell, M. and Seo, H. and Shafieloo, A. and Sharples, R. and Silber, J. and Sinigaglia, F. and Sprayberry, D. and Tan, T. and Tarl\'e, G. and Taylor, P. and Turner, W. and Ure\~na-L\'opez, L. A. and Vaisakh, R. and Valdes, F. and Valogiannis, G. and Vargas-Maga\~na, M. and Verde, L. and Walther, M. and Weaver, B. A. and Weinberg, D. H. and White, M. and Wolfson, M. and Y\`eche, C. and Yu, J. and Zaborowski, E. A. and Zarrouk, P. and Zhai, Z. and Zhang, H. and Zhao, C. and Zhao, G. B. and Zhou, R. and Zou, H.},
  collaboration = {DESI Collaboration},
  journal = {Phys. Rev. D},
  volume = {112},
  issue = {8},
  pages = {083515},
  numpages = {40},
  year = {2025},
  month = {Oct},
  publisher = {American Physical Society},
  doi = {10.1103/tr6y-kpc6},
  url = {https://link.aps.org/doi/10.1103/tr6y-kpc6}
}

@article{Chen_2019,
doi = {10.1088/1475-7516/2019/02/028},
url = {https://doi.org/10.1088/1475-7516/2019/02/028},
year = {2019},
month = {feb},
publisher = {},
volume = {2019},
number = {02},
pages = {028},
author = {Chen, Lu and Huang, Qing-Guo and Wang, Ke},
title = {Distance priors from Planck final release},
journal = {Journal of Cosmology and Astroparticle Physics}
}

@article{Copeland:2006wr,
    author = "Copeland, Edmund J. and Sami, M. and Tsujikawa, Shinji",
    title = "Dynamics of dark energy",
    eprint = "hep-th/0603057",
    archivePrefix = "arXiv",
    doi = "10.1142/S021827180600942X",
    journal = "Int. J. Mod. Phys. D",
    volume = "15",
    pages = "1753--1936",
    year = "2006"
}

@article{Weinberg:1988cp,
  title = {The cosmological constant problem},
  author = {Weinberg, Steven},
  journal = {Rev. Mod. Phys.},
  volume = {61},
  issue = {1},
  pages = {1--23},
  numpages = {0},
  year = {1989},
  month = {Jan},
  publisher = {American Physical Society},
  doi = {10.1103/RevModPhys.61.1},
  url = {https://link.aps.org/doi/10.1103/RevModPhys.61.1}
}

@article{Buchmuller:1988wx,
title = {Einstein gravity from restricted coordinate invariance},
journal = {Physics Letters B},
volume = {207},
number = {3},
pages = {292-294},
year = {1988},
issn = {0370-2693},
doi = {https://doi.org/10.1016/0370-2693(88)90577-1},
url = {https://www.sciencedirect.com/science/article/pii/0370269388905771},
author = {W. Buchmüller and N. Dragon}
}

@article{Unruh:1988in,
  title = {Unimodular theory of canonical quantum gravity},
  author = {Unruh, W. G.},
  journal = {Phys. Rev. D},
  volume = {40},
  issue = {4},
  pages = {1048--1052},
  numpages = {0},
  year = {1989},
  month = {Aug},
  publisher = {American Physical Society},
  doi = {10.1103/PhysRevD.40.1048},
  url = {https://link.aps.org/doi/10.1103/PhysRevD.40.1048}
}

@article{Jain:2012gc,
doi = {10.1088/1475-7516/2012/11/003},
url = {https://dx.doi.org/10.1088/1475-7516/2012/11/003},
year = {2012},
month = {nov},
publisher = {},
volume = {2012},
number = {11},
pages = {003},
author = {Pankaj Jain and  Atul Jaiswal and  Purnendu Karmakar and  Gopal Kashyap and  Naveen K. Singh},
title = {Cosmological implications of unimodular gravity},
journal = {Journal of Cosmology and Astroparticle Physics}
}

@article{Nojiri:2016ppu,
doi = {10.1088/0264-9381/33/12/125017},
url = {https://dx.doi.org/10.1088/0264-9381/33/12/125017},
year = {2016},
month = {may},
publisher = {IOP Publishing},
volume = {33},
number = {12},
pages = {125017},
author = {S Nojiri and S D Odintsov and V K Oikonomou},
title = {Unimodular-mimetic cosmology},
journal = {Classical and Quantum Gravity}
}

@article{Bamba:2016wjm,
author = {Bamba, Kazuharu and Odintsov, Sergei D. and Saridakis, Emmanuel N.},
title = {Inflationary cosmology in unimodular F(T) gravity},
journal = {Modern Physics Letters A},
volume = {32},
number = {21},
pages = {1750114},
year = {2017},
doi = {10.1142/S0217732317501140},
URL = { https://doi.org/10.1142/S0217732317501140},
eprint = { }
}

@article{Costantini:2022nof,
    author = "Costantini, A. and Elizalde, E.",
    title = "{A reconstruction method for anisotropic universes in unimodular F(R)-gravity}",
    eprint = "2212.02568",
    archivePrefix = "arXiv",
    primaryClass = "gr-qc",
    doi = "10.1140/epjc/s10052-022-11112-3",
    journal = "Eur. Phys. J. C",
    volume = "82",
    number = "12",
    pages = "1127",
    year = "2022"
}

@article{Cho:2014taa,
doi = {10.1088/0264-9381/32/13/135020},
url = {https://dx.doi.org/10.1088/0264-9381/32/13/135020},
year = {2015},
month = {jun},
publisher = {IOP Publishing},
volume = {32},
number = {13},
pages = {135020},
author = {Inyong Cho and Naveen K Singh},
title = {Unimodular theory of gravity and inflation},
journal = {Classical and Quantum Gravity}
}

@article{Singh:2012sx,
author = {SINGH, NAVEEN K.},
title = {UNIMODULAR CONSTRAINT ON GLOBAL SCALE INVARIANCE},
journal = {Modern Physics Letters A},
volume = {28},
number = {30},
pages = {1350130},
year = {2013},
doi = {10.1142/S0217732313501307},

URL = {https://doi.org/10.1142/S0217732313501307},
eprint = {https://doi.org/10.1142/S0217732313501307}
}

@ARTICLE{Odintsov:2016imq,
       author = "Odintsov, S.~D. and {Oikonomou}, V.~K.",
        title = "{Unimodular mimetic F(R) inflation}",
      journal = {Astrophysics and Space Science},
         year = 2016,
        month = jul,
       volume = {361},
       number = {7},
          eid = {236},
        pages = {236},
          doi = {10.1007/s10509-016-2826-9},
archivePrefix = {arXiv},
       eprint = {1602.05645},
 primaryClass = {gr-qc},
       adsurl = {https://ui.adsabs.harvard.edu/abs/2016Ap&SS.361..236O}
}

@ARTICLE{Agrawal2023,
author={Agrawal, A.S. and Mishra, B. and Moraes, P.H.R.S.},
title={Unimodular gravity traversable wormholes},
journal={European Physical Journal Plus},
year={2023},
volume={138},
number={3},
doi={10.1140/epjp/s13360-023-03872-y},
art_number={275},
note={cited By 0},
url={},
document_type={Article},
source={Scopus},
}

@article{Guin_2025,
doi = {10.1088/1475-7516/2025/11/048},
url = {https://doi.org/10.1088/1475-7516/2025/11/048},
year = {2025},
month = {nov},
publisher = {IOP Publishing},
volume = {2025},
number = {11},
pages = {048},
author = {Guin, Gopinath and Paul, Souvik and Gangopadhyay, Sunandan},
title = {Barrow holographic dark energy interacting model in the presence of radiation and matter},
journal = {Journal of Cosmology and Astroparticle Physics}
}

@article{Yang_2018,
doi = {10.1088/1475-7516/2018/09/019},
url = {https://doi.org/10.1088/1475-7516/2018/09/019},
year = {2018},
month = {sep},
publisher = {},
volume = {2018},
number = {09},
pages = {019},
author = {Yang, Weiqiang and Pan, Supriya and Valentino, Eleonora Di and Nunes, Rafael C. and Vagnozzi, Sunny and Mota, David F.},
title = {Tale of stable interacting dark energy, observational signatures, and the H0 tension},
journal = {Journal of Cosmology and Astroparticle Physics}
}

@article{DIVALENTINO_2020,
title = {Interacting dark energy in the early 2020s: A promising solution to the H0 and cosmic shear tensions},
journal = {Physics of the Dark Universe},
volume = {30},
pages = {100666},
year = {2020},
issn = {2212-6864},
doi = {https://doi.org/10.1016/j.dark.2020.100666},
url = {https://www.sciencedirect.com/science/article/pii/S2212686420300601},
author = {Eleonora {Di Valentino} and Alessandro Melchiorri and Olga Mena and Sunny Vagnozzi}
}

@article{PhysRevD.105.123506,
  title = {New tests of dark sector interactions from the full-shape galaxy power spectrum},
  author = {Nunes, Rafael C. and Vagnozzi, Sunny and Kumar, Suresh and Di Valentino, Eleonora and Mena, Olga},
  journal = {Phys. Rev. D},
  volume = {105},
  issue = {12},
  pages = {123506},
  numpages = {18},
  year = {2022},
  month = {Jun},
  publisher = {American Physical Society},
  doi = {10.1103/PhysRevD.105.123506},
  url = {https://link.aps.org/doi/10.1103/PhysRevD.105.123506}
}

@article{Wands:2012vg,
  author         = {Wands, David and De-Santiago, Jon and Wang, Yuting},
  title          = {Inhomogeneous vacuum energy},
  journal        = {Classical and Quantum Gravity},
  volume         = {29},
  pages          = {145017},
  year           = {2012},
  doi            = {10.1088/0264-9381/29/14/145017},
  eprint         = {1203.6776},
  archivePrefix  = {arXiv},
  primaryClass   = {astro-ph.CO}
}

@article{Borges:2020cbh,
  author         = {Borges, Henrique A. and Wands, David},
  title          = {Growth of structure in interacting vacuum cosmologies},
  journal        = {Physical Review D},
  volume         = {101},
  number         = {10},
  pages          = {103519},
  year           = {2020},
  doi            = {10.1103/PhysRevD.101.103519},
  eprint         = {1709.08933},
  archivePrefix  = {arXiv},
  primaryClass   = {astro-ph.CO}
}

@article{Wang:2014xca,
  author         = {Wang, Yuting and Wands, David and Zhao, Gong-Bo},
  title          = {Post-Planck constraints on interacting vacuum energy},
  journal        = {Physical Review D},
  volume         = {90},
  number         = {2},
  pages          = {023502},
  year           = {2014},
  doi            = {10.1103/PhysRevD.90.023502},
  eprint         = {1404.5706},
  archivePrefix  = {arXiv},
  primaryClass   = {astro-ph.CO}
}

@article{Wang:2015wga,
  author         = {Wang, Yuting and Zhao, Gong-Bo and Wands, David and Pogosian, Levon and Crittenden, Robert},
  title          = {Reconstruction of the dark matter--vacuum energy interaction},
  journal        = {Physical Review D},
  volume         = {92},
  pages          = {103005},
  year           = {2015},
  doi            = {10.1103/PhysRevD.92.103005},
  eprint         = {1505.01373},
  archivePrefix  = {arXiv},
  primaryClass   = {astro-ph.CO}
}

@article{Kaeonikhom:2022jyu,
  author         = {Kaeonikhom, Chayapol and Assadullahi, Hooshyar and Schewtschenko, Julia and Wands, David},
  title          = {Observational constraints on interacting vacuum energy with linear interactions},
  journal        = {Journal of Cosmology and Astroparticle Physics},
  volume         = {01},
  pages          = {042},
  year           = {2023},
  doi            = {10.1088/1475-7516/2023/01/042},
  eprint         = {2210.05363},
  archivePrefix  = {arXiv},
  primaryClass   = {astro-ph.CO}
}

@article{BOSS:2016,
    author = "Alam, Shadab and others",
    collaboration = "BOSS",
    title = "{The clustering of galaxies in the completed SDSS-III Baryon Oscillation Spectroscopic Survey: cosmological analysis of the DR12 galaxy sample}",
    eprint = "1607.03155",
    archivePrefix = "arXiv",
    primaryClass = "astro-ph.CO",
    doi = "10.1093/mnras/stx721",
    journal = "Mon. Not. Roy. Astron. Soc.",
    volume = "470",
    number = "3",
    pages = "2617--2652",
    year = "2017"
}

@article{He_2008,
doi = {10.1088/1475-7516/2008/06/010},
url = {https://doi.org/10.1088/1475-7516/2008/06/010},
year = {2008},
month = {jun},
publisher = {},
volume = {2008},
number = {06},
pages = {010},
author = {He, Jian-Hua and Wang, Bin},
title = {Effects of the interaction between dark energy and dark matter on cosmological parameters},
journal = {Journal of Cosmology and Astroparticle Physics}
}

@article{Jimenez:2003iv,
    author = "Jimenez, Raul and Verde, Licia and Treu, Tommaso and Stern, Daniel",
    title = "{Constraints on the equation of state of dark energy and the Hubble constant from stellar ages and the CMB}",
    eprint = "astro-ph/0302560",
    archivePrefix = "arXiv",
    doi = "10.1086/376595",
    journal = "Astrophys. J.",
    volume = "593",
    pages = "622--629",
    year = "2003"
}

@article{Simon:2004tf,
    author = "Simon, Joan and Verde, Licia and Jimenez, Raul",
    title = "{Constraints on the redshift dependence of the dark energy potential}",
    eprint = "astro-ph/0412269",
    archivePrefix = "arXiv",
    doi = "10.1103/PhysRevD.71.123001",
    journal = "Phys. Rev. D",
    volume = "71",
    pages = "123001",
    year = "2005"
}

@article{Stern:2009ep,
	doi = {10.1088/1475-7516/2010/02/008},

	url = {https://doi.org/10.1088%2F1475-7516%2F2010%2F02%2F008},

	year = 2010,
	month = {feb},

	publisher = {{IOP} Publishing},

	volume = {2010},

	number = {02},

	pages = {008--008},

	author = {Daniel Stern and  Raul Jimenez and  Licia Verde and  Marc Kamionkowski and  S. Adam Stanford},

	title = {Cosmic chronometers: constraining the equation of state of dark energy. I: H(z) measurements},

	journal = {Journal of Cosmology and Astroparticle Physics}
}

@ARTICLE{Moresco:2012jh,
       author = "Moresco, M. and {Cimatti}, A. and {Jimenez}, R. and {Pozzetti}, L. and {Zamorani}, G. and {Bolzonella}, M. and {Dunlop}, J. and {Lamareille}, F. and {Mignoli}, M. and {Pearce}, H. and {Rosati}, P. and {Stern}, D. and {Verde}, L. and {Zucca}, E. and {Carollo}, C.~M. and {Contini}, T. and {Kneib}, J. -P. and {Le F{\`e}vre}, O. and {Lilly}, S.~J. and {Mainieri}, V. and {Renzini}, A. and {Scodeggio}, M. and {Balestra}, I. and {Gobat}, R. and {McLure}, R. and {Bardelli}, S. and {Bongiorno}, A. and {Caputi}, K. and {Cucciati}, O. and {de la Torre}, S. and {de Ravel}, L. and {Franzetti}, P. and {Garilli}, B. and {Iovino}, A. and {Kampczyk}, P. and {Knobel}, C. and {Kova{\v{c}}}, K. and {Le Borgne}, J. -F. and {Le Brun}, V. and {Maier}, C. and {Pell{\'o}}, R. and {Peng}, Y. and {Perez-Montero}, E. and {Presotto}, V. and {Silverman}, J.~D. and {Tanaka}, M. and {Tasca}, L.~A.~M. and {Tresse}, L. and {Vergani}, D. and {Almaini}, O. and {Barnes}, L. and {Bordoloi}, R. and {Bradshaw}, E. and {Cappi}, A. and {Chuter}, R. and {Cirasuolo}, M. and {Coppa}, G. and {Diener}, C. and {Foucaud}, S. and {Hartley}, W. and {Kamionkowski}, M. and {Koekemoer}, A.~M. and {L{\'o}pez-Sanjuan}, C. and {McCracken}, H.~J. and {Nair}, P. and {Oesch}, P. and {Stanford}, A. and {Welikala}, N.",
        title = "{Improved constraints on the expansion rate of the Universe up to z $\approx$ 1.1 from the spectroscopic evolution of cosmic chronometers}",
      journal = {Journal of Cosmology and Astroparticle Physics},
         year = 2012,
        month = aug,
       volume = {2012},
       number = {8},
          eid = {006},
        pages = {006},
          doi = {10.1088/1475-7516/2012/08/006},
archivePrefix = {arXiv},
       eprint = {1201.3609},
 primaryClass = {astro-ph.CO},
       adsurl = {https://ui.adsabs.harvard.edu/abs/2012Journal of Cosmology and Astroparticle Physics...08..006M}
}

@ARTICLE{Zhang:2012mp,
       author = {{Zhang}, Cong and {Zhang}, Han and {Yuan}, Shuo and {Liu}, Siqi and {Zhang}, Tong-Jie and {Sun}, Yan-Chun},
        title = "{Four new observational H(z) data from luminous red galaxies in the Sloan Digital Sky Survey data release seven}",
      journal = {Research in Astronomy and Astrophysics},
         year = 2014,
        month = oct,
       volume = {14},
       number = {10},
          eid = {1221-1233},
        pages = {1221-1233},
          doi = {10.1088/1674-4527/14/10/002},
archivePrefix = {arXiv},
       eprint = {1207.4541},
 primaryClass = {astro-ph.CO},
       adsurl = {https://ui.adsabs.harvard.edu/abs/2014RAA....14.1221Z}
}

@ARTICLE{Moresco:2015cya,
       author = {{Moresco}, M.},
        title = "{Raising the bar: new constraints on the Hubble parameter with cosmic chronometers at z \raisebox{-0.5ex}\textasciitilde 2.}",
      journal = {Monthly Notices of the Royal Astronomical Society},
         year = 2015,
        month = jun,
       volume = {450},
        pages = {L16-L20},
          doi = {10.1093/mnrasl/slv037},
archivePrefix = {arXiv},
       eprint = {1503.01116},
 primaryClass = {astro-ph.CO},
       adsurl = {https://ui.adsabs.harvard.edu/abs/2015MNRAS.450L..16M}
}

@article{Moresco:2016mzx,
    author = "Moresco, Michele and Pozzetti, Lucia and Cimatti, Andrea and Jimenez, Raul and Maraston, Claudia and Verde, Licia and Thomas, Daniel and Citro, Annalisa and Tojeiro, Rita and Wilkinson, David",
    title = "{A 6\% measurement of the Hubble parameter at $z\sim0.45$: direct evidence of the epoch of cosmic re-acceleration}",
    eprint = "1601.01701",
    archivePrefix = "arXiv",
    primaryClass = "astro-ph.CO",
    doi = "10.1088/1475-7516/2016/05/014",
    journal = "Journal of Cosmology and Astroparticle Physics",
    volume = "05",
    pages = "014",
    year = "2016"
}

@article{Ratsimbazafy:2017vga,
    author = {Ratsimbazafy, A. L. and Loubser, S. I. and Crawford, S. M. and Cress, C. M. and Bassett, B. A. and Nichol, R. C. and V\"ais\"anen, P.},
    title = "{Age-dating Luminous Red Galaxies observed with the Southern African Large Telescope}",
    eprint = "1702.00418",
    archivePrefix = "arXiv",
    primaryClass = "astro-ph.CO",
    doi = "10.1093/mnras/stx301",
    journal = "Mon. Not. Roy. Astron. Soc.",
    volume = "467",
    number = "3",
    pages = "3239--3254",
    year = "2017"
}

@article{emcee,
   author = {{Foreman-Mackey}, D. and {Hogg}, D.~W. and {Lang}, D. and {Goodman}, J.},
    title = {emcee: The MCMC Hammer},
  journal = {PASP},
     year = 2013,
   volume = 125,
    pages = {306-312},
   eprint = {1202.3665},
      doi = {10.1086/670067}
}

@article{Borghi_2022,
doi = {10.3847/2041-8213/ac3fb2},
url = {https://doi.org/10.3847/2041-8213/ac3fb2},
year = {2022},
month = {mar},
publisher = {The American Astronomical Society},
volume = {928},
number = {1},
pages = {L4},
author = {Borghi, Nicola and Moresco, Michele and Cimatti, Andrea},
title = {Toward a Better Understanding of Cosmic Chronometers: A New Measurement of H(z) at z = 0.7},
journal = {The Astrophysical Journal Letters}
}

@article{ForemanMackey2013,
  author  = {Foreman-Mackey, Daniel and Hogg, David W. and Lang, Dustin and Goodman, Jonathan},
  title   = {emcee: The MCMC Hammer},
  journal = {Publications of the Astronomical Society of the Pacific},
  volume  = {125},
  pages   = {306--312},
  year    = {2013},
  doi     = {10.1086/670067}
}

@article{Jain:2011jc,
doi = {10.1088/1475-7516/2012/05/020},
url = {https://dx.doi.org/10.1088/1475-7516/2012/05/020},
year = {2012},
month = {may},
publisher = {},
volume = {2012},
number = {05},
pages = {020},
author = {Pankaj Jain and  Purnendu Karmakar and  Subhadip Mitra and  Sukanta Panda and  Naveen K. Singh},
title = {Testing unimodular gravity},
journal = {Journal of Cosmology and Astroparticle Physics}
}

@ARTICLE{Cedeno2021,
author={Linares Cedeño, F.X. and Nucamendi, U.},
title={Revisiting cosmological diffusion models in Unimodular Gravity and the H0 tension},
journal={Physics of the Dark Universe},
year={2021},
volume={32},
doi={10.1016/j.dark.2021.100807},
art_number={100807},
note={cited By 11},
document_type={Article},
source={Scopus},
}

@ARTICLE{Gao2014,
author={Gao, C. and Brandenberger, R.H. and Cai, Y. and Chen, P.},
title={Cosmological perturbations in unimodular gravity},
journal={Journal of Cosmology and Astroparticle Physics},
year={2014},
volume={2014},
number={9},
doi={10.1088/1475-7516/2014/09/021},
art_number={021},
note={cited By 46},
document_type={Article},
source={Scopus},
}

@Article{universe9110469,
AUTHOR = {Singh, Naveen K. and Kashyap, Gopal},
TITLE = {Unimodular Theory of Gravity in Light of the Latest Cosmological Data},
JOURNAL = {Universe},
VOLUME = {9},
YEAR = {2023},
NUMBER = {11},
ARTICLE-NUMBER = {469},
URL = {https://www.mdpi.com/2218-1997/9/11/469},
ISSN = {2218-1997},
DOI = {10.3390/universe9110469}
}

@article{PhysRevD.102.023508,
  title = {Diffusion in unimodular gravity: Analytical solutions, late-time acceleration, and cosmological constraints},
  author = {Corral, Crist\'obal and Cruz, Norman and Gonz\'alez, Esteban},
  journal = {Phys. Rev. D},
  volume = {102},
  issue = {2},
  pages = {023508},
  numpages = {17},
  year = {2020},
  month = {Jul},
  publisher = {American Physical Society},
  doi = {10.1103/PhysRevD.102.023508},
  url = {https://link.aps.org/doi/10.1103/PhysRevD.102.023508}
}

@article{cedeno_2021,
title = {Revisiting cosmological diffusion models in Unimodular Gravity and the H0 tension},
journal = {Physics of the Dark Universe},
volume = {32},
pages = {100807},
year = {2021},
issn = {2212-6864},
doi = {https://doi.org/10.1016/j.dark.2021.100807},
url = {https://www.sciencedirect.com/science/article/pii/S2212686421000388},
author = {Francisco X. {Linares Cedeño} and Ulises Nucamendi}
}

@article{PhysRevLett.118.021102,
  title = {Dark Energy from Violation of Energy Conservation},
  author = {Josset, Thibaut and Perez, Alejandro and Sudarsky, Daniel},
  journal = {Phys. Rev. Lett.},
  volume = {118},
  issue = {2},
  pages = {021102},
  numpages = {5},
  year = {2017},
  month = {Jan},
  publisher = {American Physical Society},
  doi = {10.1103/PhysRevLett.118.021102},
  url = {https://link.aps.org/doi/10.1103/PhysRevLett.118.021102}
}

@article{Das_2023,
doi = {10.1088/1475-7516/2023/10/047},
url = {https://doi.org/10.1088/1475-7516/2023/10/047},
year = {2023},
month = {oct},
publisher = {IOP Publishing},
volume = {2023},
number = {10},
pages = {047},
author = {Das, Santanu and Nasiri, Arad and Yazdi, Yasaman K.},
title = {Aspects of Everpresent $\Lambda$. Part I. A fluctuating cosmological constant from spacetime discreteness},
journal = {Journal of Cosmology and Astroparticle Physics}
}

\end{document}